\documentclass[
reprint,
superscriptaddress,
 amsmath,amssymb,
prb,
floatfix,
]{revtex4-2}
\usepackage{dcolumn}
\usepackage{bm}
\usepackage{turnstile}

\usepackage{graphicx}
\usepackage{amsmath,amssymb}
\usepackage{url}

\usepackage{epsfig}
\usepackage{epstopdf}

\begin{document}

\title{
Chirality of Bloch domain walls in exchange biased CoO/Co bilayer seen by waveguide-enhanced neutron spin-flip scattering
}
\author{Yu.~N.~Khaydukov}
\affiliation{Max-Planck-Institut f\"ur Festk\"orperforschung, Heisenbergstra\ss e 1, D-70569 Stuttgart, Germany}
\affiliation{Max Planck Society Outstation at the Heinz Maier-Leibnitz Zentrum (MLZ), D-85748 Garching, Germany}
\affiliation{Skobeltsyn Institute of Nuclear Physics, Moscow State University, Moscow 119991, Russia}

\author{D.~Lenk}
\affiliation{Institut f\"ur Physik, Universit\"at Augsburg, D-86158 Augsburg, Germany}
\author{V.~Zdravkov}
\affiliation{Institut f\"ur Physik, Universit\"at Augsburg, D-86158 Augsburg, Germany}
\affiliation{Institute of Electronic Engineering and Nanotechnologies ASM, MD2028 Kishinev, Moldova}
\author{R.~Morari}
\affiliation{Institut f\"ur Physik, Universit\"at Augsburg, D-86158 Augsburg, Germany}
\affiliation{Institute of Electronic Engineering and Nanotechnologies ASM, MD2028 Kishinev, Moldova}
\affiliation{Topological Quantum Phenomena in Superconducting Systems Lab  (TQPSS),MIPT, Dolgoprudny, Russia}
\author{T.~Keller}
\affiliation{Max-Planck-Institut f\"ur Festk\"orperforschung, Heisenbergstra\ss e 1, D-70569 Stuttgart, Germany}
\affiliation{Max Planck Society Outstation at the Heinz Maier-Leibnitz Zentrum (MLZ), D-85748 Garching, Germany}
\author{A.~S.~Sidorenko}
\affiliation{Institute of Electronic Engineering and Nanotechnologies ASM, MD2028 Kishinev, Moldova}
\affiliation{Topological Quantum Phenomena in Superconducting Systems Lab  (TQPSS),MIPT, Dolgoprudny, Russia}
\author{L.~R.~Tagirov}
\affiliation{Institut f\"ur Physik, Universit\"at Augsburg, D-86158 Augsburg, Germany}
\affiliation{E.K. Zavoisky Physical-Technical Institute of RAS, 420029 Kazan, Russia}
\author{R.~Tidecks}
\affiliation{Institut f\"ur Physik, Universit\"at Augsburg, D-86158 Augsburg, Germany}
\author{S.~Horn}
\affiliation{Institut f\"ur Physik, Universit\"at Augsburg, D-86158 Augsburg, Germany}
\author{B.~Keimer}
\affiliation{Max-Planck-Institut f\"ur Festk\"orperforschung, Heisenbergstra\ss e 1, D-70569 Stuttgart, Germany}
\date{\today}

\begin{abstract}
Magnetic state of exchanged biased CoO(20nm)/Co($d_F$) bilayer ($d_F$=5-20nm) was studied by means of polarized neutron reflectometry. By spacing of CoO/Co bilayer and Al$_2$O$_3$ substrate with Nb(20nm) layer we created waveguide structure which allowed us to significantly enhance intensity of spin-flip (SF) scattering in the position of optical resonances. For the trained sample with thinnest Co(5nm) we detected strong SF scattering at the resonance position (up to 30\% of incoming intensity) speaking about high non-collinearity of the system. As $d_F$ increases, the intensity of SF scattering linearly decreases. At the same time we observed asymmetry of up-down and down-up  scattering channels at the resonance positions. We attribute this asymmetry to the Zeeman splitting of neutrons energies with different initial polarization taking place in high external field. Analysis, however, shows that the applied in the PNR experiment external field is not enough to quantitatively explain the observed asymmetry for the samples with $d_F > $ 5nm and we have to postulate presence of additional magnetic field produced by sample. We attribute this additional field to the stray field produced by chiral Bloch domain walls. The chirality of the domain walls can be explained by Dzyaloshinskii-Moriya interaction arising at the CoO/Co interface. Our results can be useful for designing of spintronic devices using exchange bias effect.
\end{abstract}

\maketitle

 \section{Introduction}
The exchange bias (EB) effect arises on the interface of antiferromagnetic (AF) and ferromagnetic (F) magnetic phases and leads to the shift of the hysteresis loop on value of $H_{eb}$ varying from several Oersteds to kOe (see review \cite{RaduZabel08}). The effect was discovered 60 years ago, well-studied to the date \cite{RaduZabel08,LiuRev12} and is utilized in creation of spin-valves including superconducting devices \cite{Tagirov18,Kushnir18,LenkPRB17,LeksinNanoRes16,LenkBJN16,StamopoulosAPL14,BanerjeeNatCom14,FlokstraPRB15,
ZdravkovAPL13,ZdravkovPRB13,LeksinPRB12,LeksinPRL12}.

One of the widely used methods in studying of EB is Polarized Neutron Reflectometry  (PNR) \cite{Fitzsimmons2000,LeightonPRL01,LeePRB02,RaduApplPhys02,RaduPhysB03,RaduPRB03,RaduJOP05,
PaulJOP06,PaulPRB06,PaulJOP07,PaulJAP10,Cortie12,Demeter12,PaulPRB13,KhaydukovLowT17,Kim19,Chen19}. In this method reflectivities R$^{\mu \nu}$  are measured as a function of momentum transfer $Q = 4\pi\sin\theta/\lambda$. Here $\theta$ and $\lambda$ are incident angle and neutron wavelength, indices $\mu, \nu$  take values + or - and denote the neutron spin projection parallel to external field $H$ before and after the scattering process, respectively. The non-spin-flip (NSF) reflectivities $R^{++}$ and $R^{--}$ are sensitive to nuclear scattering length density (SLD) of the system $\rho(z)$  and in-plane component of the magnetization parallel to external field ($M_{||}(z)$): $R^{\pm \pm} \sim (\rho(z) \pm M_{||}(z))$. For analysis of $M_{||}(z)$ it is convenient to use spin asymmetry $S = (R^{++}-R^{--})/(R^{++}+R^{--})$, which is proportional to $M_{||}(z)$. The non-collinear component of the magnetization $M_{\perp}(z)$ in turn causes spin-flip (SF) scattering: $R^{\pm \mp} \sim M_{\perp}(z)^2 $. To obtain magnetic depth profiles $M_{||}(z)$ and $M_{\perp}(z)$
one needs to fit simultaneously theoretical curves in all spin channels to experimental ones by varying parameters of the nuclear and magnetic potential. This is a fairly time-consuming procedure that requires initial knowledge of both the composition of sample and  instrument itself. Also, accurate and adequate recovery of nuclear and magnetic parameters requires involvement of complementary techniques.  There is, however, a simple model-free approach which allows one for quick analysis of the magnetic configuration. In this approach \cite{LeePRB02} one may write

\begin{equation}\label{SA}
S(\alpha)/S_{sat}(0)=cos \alpha
\end{equation}

\begin{equation}\label{SF}
R^{SF}(\alpha)/R^{SF}_{sat}(90^\circ) = sin^2 \alpha ,
\end{equation}

where $\alpha$ is the angle between $\textbf{M}$ and $\textbf{H}$ (Fig.\ref{Fig1}), $S_{sat}(0)$ is the spin asymmetry in the saturation state with $\alpha$ = 0$^\circ$ and $R^{SF}_{sat}(90^\circ)$ is the maximum possible SF reflectivity at fully saturated state with $\alpha$ = 90$^\circ$ (Fig.\ref{Fig1}).  Expressions \eqref{SA}, \eqref{SF} are valid for all $Q$ values of a single F layer system being in a single domain state. For the multidomain state two limiting cases can be considered, depending on the relation between typical in-plane domain size $\xi$ and neutron coherence length $l_{coh}$.  "Big domains" ($\xi > l_{coh}$) will scatter in specular direction ($\theta_2=\theta_1$) and resulting reflectivities will be incoherently summed for all domains. For this case statistical averaging of \eqref{SA}, \eqref{SF} over all domains  allows one to obtain information about domain dispersion $\sigma^2$:

\begin{equation}\label{sigma}
 \begin{split}
\sigma^2 = <cos^2 \alpha> - <cos \alpha>^2 =
\\         1-<sin^2 \alpha>-<cos \alpha>^2
 \end{split}
\end{equation}

One can distinguish cases of single domain ($\sigma^2$=0), domains which are collinear to $H$ ($\sigma^2$=1) or isotropic distributed domains ($\sigma^2$=0.5). For the case of "small" domains with $\xi < l_{coh}$ neutrons will be scattered  both in specular and off-specular ($\theta_2 \neq \theta_1$) directions. Specular reflectivities in this case will depend on the in-plane averaged vector of magnetization with amplitude less than the saturation value and, hence, expressions \eqref{SA},\eqref{SF} are not valid anymore. For this case one needs to fit simultaneously specular and off-specular data using existing approaches \cite{Zabel07,Toperverg15}.

Thus it can be seen that the spin-flip scattering is an important channel which gives an unique information about non-collinear part of magnetization and hence processes taking place in EB system during magnetization reversal.  One needs, however to notice, that SF scattering, being of purely magnetic origin, is a fairly weak channel. To increase intensity of SF scattering one may e.g. increase number of investigated AF/F bilayers \cite{RaduApplPhys02,PaulPRB06,PaulJOP07,PaulJAP10,PaulPRB13}. Another approach is to increase time spent by a neutron in the vicinity of the investigated AF/F bilayer. This is done by placing of investigated bilayer between two layers with relatively high SLD. The formed thus optical potential allows to trap neutrons at certain $Q$-value below critical edge $Q_{crit}$ of total reflection. This bond state leads to increased time which neutron spend in the vicinity of investigated F layer and hence increases significantly the probability of SF scattering.  For the study of EB the waveguide approach was first used by Radu et al \cite{RaduPRB03}. To construct a waveguide structure two additional and relatively thick layers of Ti(200nm) and Cu(100nm) were deposited on the sapphire substrate prior to deposition of Co(20nm)/CoO(2.5nm) bilayer.

We propose another design for the waveguide structure which utilizes high neutron scattering ability of oxygen atoms. In this approach an oxide substrate such as MgO ($\rho = 6.0 \times 10^{-4} $nm$^{-2}$) or Al$_2$O$_3$ ($\rho = 5.5 \times 10^{-4}$ nm$^{-2}$) is used as a bottom high-SLD layer. As an upper high-SLD layer an oxide AF layer itself can be used. Since a waveguide requires certain thickness ($\sim$ 10nm) to form a neutron standing wave, we also used a Nb(20nm) spacer in the design. Niobium itself is a superconducting material with bulk transition temperature $T_C$ = 10 K, so the same structures can be used to study proximity effects at  superconducting/ferromagnet interface. Thus the proposed waveguide design CoO(20nm)/Co($d_F$)/Nb(20nm)//Al$_2$O$_3$ allows one to decrease both the number of deposited layers and the total thickness of the structure improving thus the quality of the structure.

\section{Samples preparation and experimental techniques}
\begin{figure}[htb]
\centering
\includegraphics[width=\columnwidth]{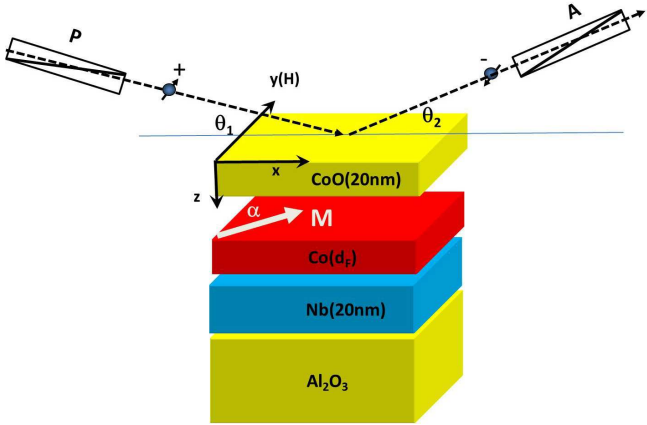}
\caption{
Sketch of samples and PNR experiment. Polarizer and analyzer are shown by rectangulars depicted as $P$ and $A$. External field at the sample position is applied along $y$-axis. Averaged magnetization of Co $M$ makes angle $\alpha$  with $H$.
 }
\label{Fig1}
\end{figure}

Samples with nominal composition CoO(20nm)/Co($d_F$)/Nb(20nm) were prepared in the magnetron sputtering machine Leybold Z-400 on Al$_2$O$_3$  substrates with (1$\bar{1}$02) surface orientation (Fig. \ref{Fig1}). For the deposition we used targets of niobium and cobalts with purity of 99.99\%. Sputtering was done in undiluted Argon 8$\times10^{-3}$  mbar atmosphere, for pure materials (Nb/Co structures), and in Ar(83\%) + O$_2$(17\%) mixture of 9.6$\times10^{-3}$  mbar total pressure for an antiferromagnetic CoO$_x$ film. The purity of gases were 99.999\% for Ar and O$_2$. A residual pressure in the chamber was about 1.5$\times10^{-6}$ mbar. In order to remove contaminations, such as absorbed gases and oxides, the targets and substrates were pre-sputtered for 3-5 minutes before deposition of the structures. The deposition rate of the layers was: 4.5nm/s for Nb, 0.1nm/s for Co, and 0.08nm/s for CoO$_x$. For the study of $d_F$ dependence of EB in our systems we have prepared the wedge structure with $d_F$  covering range from 0 to 20nm with step $\delta d_F$ = 1.6 nm. These samples were used to measure hysteresis loops using Superconducting Quantum Interference Device (SQUID). For the neutron study we have prepared four 10$\times$10mm$^2$ samples with $d_F$ = 5 nm(s05),10 nm(s10),15 nm(s15) and 20 nm(s20).

The neutron experiment was conducted at the monochromatic ($\lambda$ =4.3\AA$\pm$2\%) reflectometer NREX (research reactor FRM-2, Garching, Germany) with horizontal alignment of the sample (Fig. \ref{Fig1}). In vertical direction the beam was collimated to divergence of $\delta \theta_x $ = 0.25 mrad. In horizontal direction divergence was relaxed to $\delta \theta_y \sim$ 30 mrad in order to increase intensity of incoming beam. From these values we can estimate neutron coherence lengths in x and y directions \cite{Zabel07} as $l_{coh,x} \sim 10 \mu$m and $l_{coh,y} \sim 1$nm. Polarization of incoming beam was performed by passing through a 0.3 mm thick Fe/Si supermirror with efficiency P$_P$=99.99\%. External magnetic field was applied parallel to the surface and normal to the scattering plane. Polarization of the beam scattered under an angle $\theta_2$ was analyzed by similar to polarizer device with efficiency P$_A$=99.6\%. To change the polarization before and after the sample two adiabatic spin-flippers with efficiencies close to 100\% were used.

The measurements were performed at $T$=13K which is far below the blocking temperature of CoO $T_B$=180K and slightly above $T_C$ of bulk Nb. In the PNR experiment we used two protocols. In the first protocol (protocol A) sample was cooled down in maximum possible at NREX magnetic field of H$_{max}$ = 4.5kOe and trained three times prior to the PNR measurements. In order to reach $H<$0 region we used following \cite{RaduApplPhys02} cooling of the sample in negative magnetic field. To create fully non-collinear magnetic state with $\alpha$ = 90$^\circ$ we used following procedure (protocol B). First the sample was cooled in H$_{max}$ to $T$=13K. Then the field was released to zero and the sample was rotated along $z$-axis on 90 degree. After this  small magnetic field was applied and neutron reflectivity curves were measured.  The measured experimental reflectivity curves were then compared to the model as it is described in our prior works \cite{KhaydukovPRB14,KhaydukovPRB17,KhaydukovPRB19} to extract nuclear and magnetic depth profiles.

 \section{Experiment}
\begin{figure}[htb]
\centering
\includegraphics[width=\columnwidth]{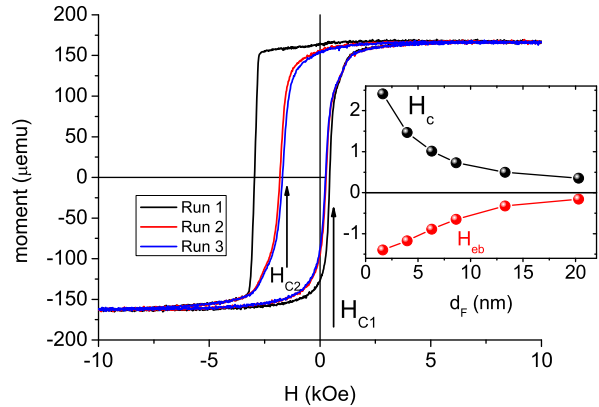}
\caption{
The hysteresis loops of the sample with $d_F$=5.0 nm measured at 13K after cooling in 4.5kOe. Vertical arrows show coercivity fields in positive and negative magnetic fields. The inset shows thickness dependence of coercive field $H_c$=($|H_{C1}|+|H_{C2}|$)/2 and exchange bias field $H_{eb}$.
 }
\label{Fig2}
\end{figure}

 Figure \ref{Fig2} shows three successive hysteresis loops taken after cooling s05 sample in $H$ = 10kOe to $T$=13K. The virgin loop shows the sharp transitions at coercive fields $H_{C1}$=-3kOe and $H_{C2}$ = +0.4kOe so that exchange bias field for the virgin sample $H_{eb}=(H_{C1}+H_{C2})/2$ = -1.25kOe. Following field sweeping leads to the gradual decrease of $H_{eb}$ (training effect) which stabilizes at $H_{eb}$ = -0.7kOe after three loops. Inset to Fig. \ref{Fig2} shows the thickness dependence of coercive and exchange bias fields in the trained state. One can notice typical for EB $H_{eb} \sim 1/d_F$ behavior which is the proof of the interfacial nature of the EB effect. The maximum EB field we measured on the thinnest sample with $d_F$ = 1.6nm is $H_{eb}$ = -1.4kOe which is one of the strongest effect observed so far (see Table 3.1 in  \cite{RaduZabel08} for comparison).

\begin{figure*}[htb]
\centering
\includegraphics[width=2\columnwidth]{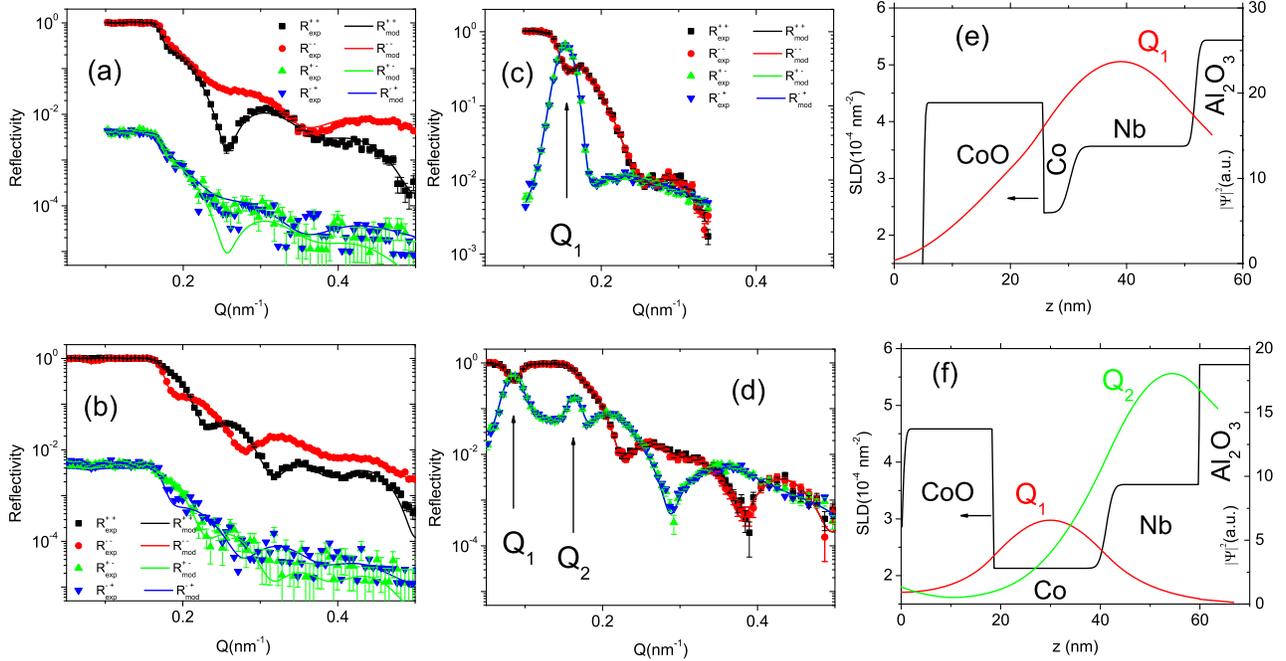}
\caption{
Experimental (dots) reflectivity curves measured in saturated state, i.e. protocol A in $H=H_{max}$ (a- $d_F$=5nm, b - $d_F$=20nm) and in fully non-collinear state, i.e. protocol B in $H$=5Oe (c- $d_F$=5nm, d - $d_F$=20nm). Solid lines in (a)-(d) corresponds to best-fit model curves calculated for the SLD depth profiles depicted in (e) and (f) for $d_F$=5nm and $d_F$=20nm. Vertical arrows in (c) and (d) show position of the resonance(s). The depth profile of neutron density at the resonances are shown in (e) and (f) by red and green lines.
}
\label{Fig3}
\end{figure*}

Figures \ref{Fig3}a and \ref{Fig3}b show the spin-polarized reflectivity curves measured on samples s05 and s20 in the saturated state. Above the critical edge $Q > Q_{crit}$ = 0.16 nm$^{-1}$ one can see strong spin asymmetry of NSF reflectivities. The SF reflectivities in turn are almost zero except a less than 1\% background caused mostly by non-ideal efficiency of the spin-analyzer. For both samples we were able to reproduce experimental reflectivities with the SLD depth profiles depicted in Fig. \ref{Fig3}e,f and with magnetization of Co within 5\% error bar close to bulk magnetization of Co 4$\pi$M = 18kG.

Figures \ref{Fig3}c and \ref{Fig3}d show reflectivity curves for samples s05 and s20 in fully non-collinear state (protocol B). In this case neutron reflectivities are characterized by zero spin asymmetry of NSF channels and strong spin-flip scattering with one or two waveguide peaks for thin ($d_F <$ 15nm) and thick ($d_F > $ 15nm) samples. Using the fit-obtained SLD depth profiles we calculated neutron density depth profiles $|\Psi(z)|^2$  at the position of resonances (Fig.\ref{Fig3}e,f).
Our calculations show that for samples with thin Co layer a neutron standing wave centered in the middle of Co/Nb bilayer is formed  at $Q$=$Q_1$.  For thick samples resonances at $Q_1$ and $Q_2$ corresponds to neutron standing waves centered in Co and Nb layers respectively. It is interesting to note that despite of 4-times difference in thickness of Co layer of s05 and s20 samples the spin-flip reflectivities at  $Q_1$ resonance are quite similar. Reason of this is a higher amplitude of $|\Psi(z)|^2$ at the position of Co layer, that is, efficiency of waveguide enhancement of s05 sample is much higher than of s20 sample.

\begin{figure*}[htb]
\centering
\includegraphics[width=2\columnwidth]{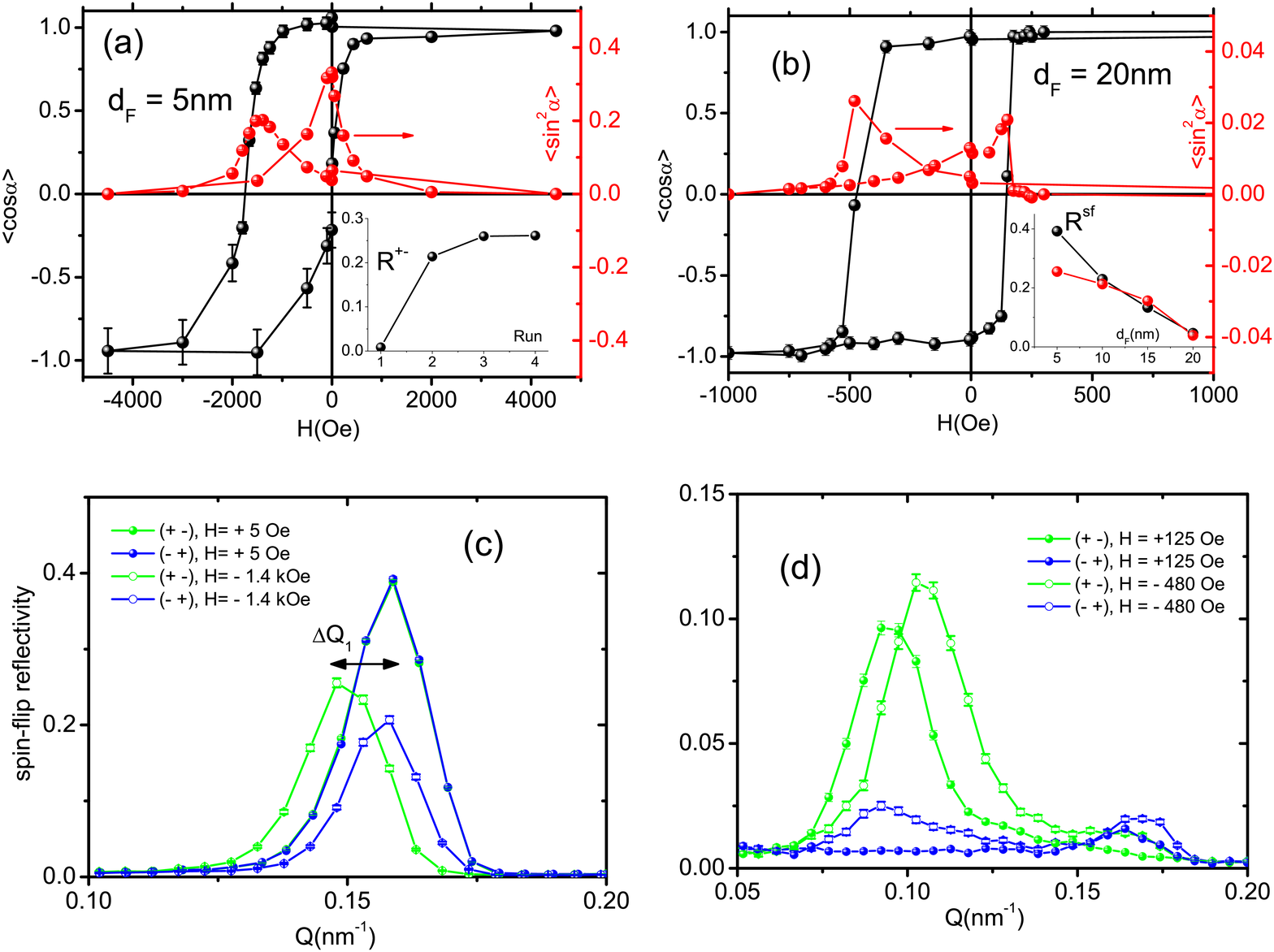}
\caption{
The field dependence of the normalized spin asymmetry and spin-flip scattering for the sample s05 (a) and s20 (b). Note the one order difference of $\langle\sin^2\alpha\rangle$  in (a) and (b). Inset in (a) shows the amplitude of spin-flip peak at $Q_1$ resonance around $H_{C1}$ versus number of field cycles.  Inset in (b) shows amplitude of spin-flip peak around $H_{C1}$(black) and $H_{C2}$(red) as a function of $d_F$.  Graphs (c) and (d) shows spin-flip reflectivities measured at left and right reversal points on sample s05 and s20. Horizontal line in (c) depicts size of peaks splitting $\delta Q_1$.
}
\label{Fig4}
\end{figure*}

Figures \ref{Fig4}a and \ref{Fig4}b shows the field dependence of the averaged cosine $\langle\cos\alpha\rangle$ (black) and squared-sine $\langle\sin^2\alpha\rangle$ (red) for samples s05 and s20 calculated from \eqref{SA}, \eqref{SF} using the data of Fig.\ref{Fig3} for normalization. To increase statistical accuracy for calculation of $\langle\cos\alpha\rangle$ we integrated data in the region of $Q>Q_{crit}$ where spin asymmetry is constant-sign function. For calculation of $\langle\sin^2\alpha\rangle$ we used spin-flip data integrated in the vicinity of $Q_1$ peak for up-down and down-up channels. From this analysis it follows that magnetization reversal of sample s05 with thinnest Co layer is characterized by a strongly non-collinear state. The amplitude of spin-flip scattering reaches value of 20\% and 30\% in the vicinity of $H_{C1}$ and $H_{C2}$. One needs to say that the high intensity of spin-flip scattering appears as a result of training (see inset in {Fig4}a).

Figures \ref{Fig4}c,d show the SF reflectivities at the reversal points for samples s05 and s20. One can see that in contrast to the artificially created non-collinear state (Fig. \ref{Fig3}c,d) the spin-flip $R^{+-}$ and $R^{-+}$ reflectivities are strongly different: for s05 sample one can observe $dQ_1$=8$\times$10$^{-3}$ nm$^{-1}$ shift of resonances between $R^{+-}$ and $R^{-+}$ while for s20 intensity of $R^{-+}$  at $Q_1$ is much less comparing to amplitude of $R^{+-}$. The only one exception are spin-flip reflectivities of s05 sample at $H$=5Oe where both curves look the same. All the rest samples have shown asymmetrical spin-flip scattering at all points where SF scattering was above the background limit.

 \section{Discussions and Conclusion}
In this work we systematically studied peculiarities of magnetization reversal of exchange biased CoO(20nm)/Co($d_F$) bilayer ($d_F$=5-20nm) using waveguide-enhanced neutron spin-flip scattering. Our study showed that trained samples are characterized by hardly erasable spin-flip scattering.  We relate this SF scattering to the existence of domain state seen before in CoO/Co systems \cite{WelpJAP03,BremsJAP08} using various microscopic methods. At the same time simultaneous analysis of spin asymmetry and spin-flip channels allows us to get more information on  mechanism of reversal and its $d_F$-dependence. So the sample s05 with  thin Co(5nm) is characterized by a strong spin-flip scattering reaching a level of 20-25\% near the reversal points. Using the expression \eqref{sigma} we can roughly estimate $\sigma^2 \sim$ 0.5, which means that samples with thin Co are remagnetized via domains rotation.  Increasing of $d_F$ leads to the linear decrease of spin-flip intensity which reaches the value of 2-3\% for $d_F$=20nm. Such a small intensity corresponds to the situation of collinear domains ($\sigma^2 \sim $ 1), and the remagnetization goes through the movement of the domain wall.
 
\begin{figure*}[htb]
\centering
\includegraphics[width=2\columnwidth]{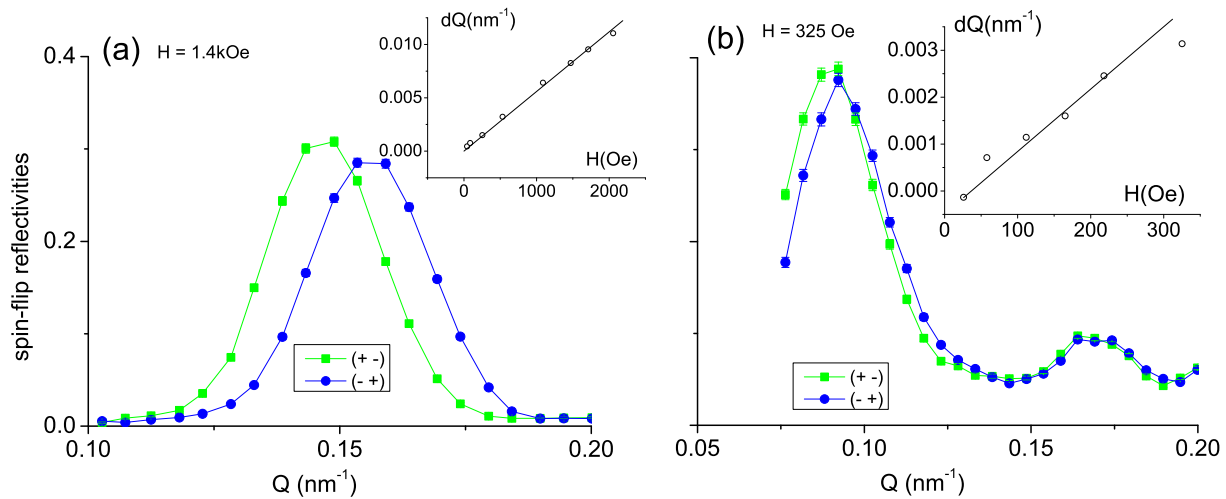}
\caption{
The SF curves measured on virgin sample s05 in  magnetic field $H$=1.4kOe (a) and virgin sample s20 in  magnetic field $H$ = 325Oe applied normally to the direction of cooling field. Insets show $H$ dependence of the peaks splitting.
}
\label{Fig5}
\end{figure*}

 An interesting feature we observed in our experiment is non-equality of up-down and down-up scattering channels. Such an asymmetry of the spin-flip signal is rather exotic phenomenon taking place under certain conditions in elastic experiment on helimagnets \cite{Schwink75,Aksenov07,Fraerman07,Grigoriev08,Tarnavich14,Tarnavich17} or inelastic experiments by magnons \cite{Gukasov99} in bulk systems. There is, however, a simpler explanation related to Zeeman splitting of the neutron energy with different polarization in a large external field \cite{felcher95,Aksenov01,Kozhevnikov12,Maranville16,Kozhevnikov17,Kozhevnikov18}. Noteworthy that a similar splitting of SF waveguide peaks observed earlier  for CoO(2.5nm)/Co(20nm) EB bilayer \cite{RaduPhysB03} was explained by the same effect.  Having a strong unidirectional anisotropy in our samples we can perform a simple experimental test whether the splitting indeed is related to the external field rather than to internal magnetic state. To do so, we measured SF reflectivities of virgin samples at fully non-collinear state (protocol B) in different magnetic fields (Fig. \ref{Fig5}) not exceeding $H_{EB}$. For these measurements we expect that internal magnetic state of Co layer remains untouched and only external field is a variable parameter. Fig. \ref{Fig5}a,b show the SF reflectivities measured in magnetic field $H$=1.4kOe for sample s05 and $H$ = 300Oe for sample s20. The experiment showed presence of peaks splitting with linear dependence on external field, $\delta Q_1 = cH$. For samples s05, s15 and s20 we obtained slopes c=5.4, 6.3 and 10.5 (units $10^{-6}$ nm$^{-1}$Oe$^{-1}$). These numbers allow us to check that for s05 the splitting of SF peaks can be fully explained by effect of external field (this can be also seen by a simple comparison of Fig.\ref{Fig4}c and Fig.\ref{Fig5}a). For the remaining samples, the situation is radically different.  So for s10 sample measurement at practical remanence ($H$=5Oe) showed splitting of the first resonant peak $\delta Q_1$=0.02nm$^{-1}$. Such a splitting would correspond to an applied field of $\sim 3$kOe. For the samples s15 and s20 we have not observed $R^{+-}$ peak at $Q_1$ position at all. If we propose that position of $R^{+-}$ peak is shifted  below possible minimum $Q$=0.05nm$^{-1}$, this gives us  estimation of $\delta Q_1 >$ 0.03nm$^{-1}$ corresponding to applied field $\sim$ 5kOe.

Thus, for samples with $d_F>$5nm the splitting of the resonance peaks can be explained by an external field significantly exceeding the applied one in experiment. Following Radu et al. \cite{RaduPhysB03} we can propose that additional component of $H$ is generated by the stray field from domain walls. Indeed, thin magnetic films due to the strong shape anisotropy  are featured with Neel domain walls when magnetization rotates in-plane.  At increase of the film thickness the shape anisotropy energy decreases and Bloch domain walls which rotate out of plane becomes energetically preferable. However, normally the number of left- and right turning Bloch walls should be the same, leading to the compensation of the stray field. Thus  we need to state the presence of helicity of the Bloch type domain walls in our samples. Chiral domain walls were observed recently in  systems with strong perpendicular anisotropy \cite{Chen13,Shahbazi18} and attributed to Dzyaloshinskii-Moriya Interaction (DMI). The DMI was also observed in EB system \cite{Ma17} so this interaction can generate  chirality of domain walls in our system as well.

In conclusion, we studied dependence of magnetic state of exchange biased CoO(20nm)/Co($d_F$) bilayer ($d_F$=5$\div$20nm) on thickness of Co layer by polarized neutron reflectometry. By spacing of CoO/Co bilayer and Al$_2$O$_3$ substrate with Nb(20nm) layer we created waveguide structure which allowed us to increase intensity of spin-flip (SF) scattering by an order. Our investigation showed that intensity of SF scattering of a trained sample strongly depends on $d_F$: for thinnest Co(5nm) we have observed strong SF scattering reaching value of 20-30\% of incident intensity, speaking thus about strong non-collinearity of magnetic state. By increasing of $d_F$ intensity of SF scattering linearly decreases reaching 2-3\% for $d_F$=20nm. At the same time we observed asymmetry of up-down and down-up  scattering channels at the resonance positions. We attribute this asymmetry to the Zeeman splitting of neutrons energies with different initial polarization taking place in high external field. Analysis, however, shows that the applied in the PNR experiment external field is not enough to quantitatively explain the observed asymmetry for the samples with $d_F > $ 5nm and we have to postulate presence of additional magnetic field produced by the sample. We attribute this additional field to the stray field produced by chiral Bloch domain walls. The chirality of the domain walls can be explained by Dzyaloshinskii-Moriya interaction arising at the CoO/Co interface. Our results can be useful for designing of spintronic devices using exchange bias effect.

\begin{acknowledgments}
We would like to thank Thomas Saerback, Boris Toperverg, Viktor Bodnarchuk, Oleg Udalov, Sergey Kozhevnikov, Michael Fitzsimmons, and Anton Devishvili  for the fruitful discussions. We would also like to express our gratitude to Franz Tralmer for his technical support. This work is based upon experiments performed at the NREX instrument operated by Max-Planck Society at the Heinz Maier-Leibnitz Zentrum (MLZ), Garching, Germany and supported by  German Research Foundation (Deutsche Forschungsgemeinschaft, DFG, Project No. 107745057 - TRR80).  AS and RM would like to thank the support of the "SPINTECH" project of the HORIZON-2020 TWINNING program (2018-2020).
\end{acknowledgments}

\bibliography{CoEB_Refs}

\begin{thebibliography}{54}%
\makeatletter
\providecommand \@ifxundefined [1]{%
 \@ifx{#1\undefined}
}%
\providecommand \@ifnum [1]{%
 \ifnum #1\expandafter \@firstoftwo
 \else \expandafter \@secondoftwo
 \fi
}%
\providecommand \@ifx [1]{%
 \ifx #1\expandafter \@firstoftwo
 \else \expandafter \@secondoftwo
 \fi
}%
\providecommand \natexlab [1]{#1}%
\providecommand \enquote  [1]{``#1''}%
\providecommand \bibnamefont  [1]{#1}%
\providecommand \bibfnamefont [1]{#1}%
\providecommand \citenamefont [1]{#1}%
\providecommand \href@noop [0]{\@secondoftwo}%
\providecommand \href [0]{\begingroup \@sanitize@url \@href}%
\providecommand \@href[1]{\@@startlink{#1}\@@href}%
\providecommand \@@href[1]{\endgroup#1\@@endlink}%
\providecommand \@sanitize@url [0]{\catcode `\\12\catcode `\$12\catcode
  `\&12\catcode `\#12\catcode `\^12\catcode `\_12\catcode `\%12\relax}%
\providecommand \@@startlink[1]{}%
\providecommand \@@endlink[0]{}%
\providecommand \url  [0]{\begingroup\@sanitize@url \@url }%
\providecommand \@url [1]{\endgroup\@href {#1}{\urlprefix }}%
\providecommand \urlprefix  [0]{URL }%
\providecommand \Eprint [0]{\href }%
\providecommand \doibase [0]{https://doi.org/}%
\providecommand \selectlanguage [0]{\@gobble}%
\providecommand \bibinfo  [0]{\@secondoftwo}%
\providecommand \bibfield  [0]{\@secondoftwo}%
\providecommand \translation [1]{[#1]}%
\providecommand \BibitemOpen [0]{}%
\providecommand \bibitemStop [0]{}%
\providecommand \bibitemNoStop [0]{.\EOS\space}%
\providecommand \EOS [0]{\spacefactor3000\relax}%
\providecommand \BibitemShut  [1]{\csname bibitem#1\endcsname}%
\let\auto@bib@innerbib\@empty
\bibitem [{\citenamefont {Radu}\ and\ \citenamefont
  {Zabel}(2008)}]{RaduZabel08}%
  \BibitemOpen
  \bibfield  {author} {\bibinfo {author} {\bibfnamefont {F.}~\bibnamefont
  {Radu}}\ and\ \bibinfo {author} {\bibfnamefont {H.}~\bibnamefont {Zabel}},\
  }\bibinfo {title} {Exchange bias effect of ferro-/antiferromagnetic
  heterostructures},\ in\ \href {https://doi.org/10.1007/978-3-540-73462-8_3}
  {\emph {\bibinfo {booktitle} {Magnetic Heterostructures: Advances and
  Perspectives in Spinstructures and Spintransport}}},\ \bibinfo {editor}
  {edited by\ \bibinfo {editor} {\bibfnamefont {H.}~\bibnamefont {Zabel}}\ and\
  \bibinfo {editor} {\bibfnamefont {S.~D.}\ \bibnamefont {Bader}}}\ (\bibinfo
  {publisher} {Springer Berlin Heidelberg},\ \bibinfo {address} {Berlin,
  Heidelberg},\ \bibinfo {year} {2008})\ pp.\ \bibinfo {pages}
  {97--184}\BibitemShut {NoStop}%
\bibitem [{\citenamefont {Liu}(2012)}]{LiuRev12}%
  \BibitemOpen
  \bibfield  {author} {\bibinfo {author} {\bibfnamefont {Z.}~\bibnamefont
  {Liu}},\ }\bibfield  {title} {\bibinfo {title} {Recent advances in exchange
  bias of layered magnetic {FM}/{AFM} systems},\ }\href
  {https://doi.org/10.1007/s11433-012-4963-7} {\bibfield  {journal} {\bibinfo
  {journal} {Science China Physics, Mechanics and Astronomy}\ }\textbf
  {\bibinfo {volume} {56}},\ \bibinfo {pages} {61} (\bibinfo {year}
  {2012})}\BibitemShut {NoStop}%
\bibitem [{\citenamefont {Tagirov}\ \emph {et~al.}(2018)\citenamefont
  {Tagirov}, \citenamefont {Kupriyanov}, \citenamefont {Kushnir},\ and\
  \citenamefont {Sidorenko}}]{Tagirov18}%
  \BibitemOpen
  \bibfield  {author} {\bibinfo {author} {\bibfnamefont {L.~R.}\ \bibnamefont
  {Tagirov}}, \bibinfo {author} {\bibfnamefont {M.~Y.}\ \bibnamefont
  {Kupriyanov}}, \bibinfo {author} {\bibfnamefont {V.~N.}\ \bibnamefont
  {Kushnir}},\ and\ \bibinfo {author} {\bibfnamefont {A.}~\bibnamefont
  {Sidorenko}},\ }\bibfield  {title} {\bibinfo {title} {Superconducting triplet
  proximity and josephson spin valves},\ }in\ \href
  {https://doi.org/10.1007/978-3-319-90481-8_2} {\emph {\bibinfo {booktitle}
  {{NanoScience} and Technology}}}\ (\bibinfo  {publisher} {Springer
  International Publishing},\ \bibinfo {year} {2018})\ pp.\ \bibinfo {pages}
  {31--47}\BibitemShut {NoStop}%
\bibitem [{\citenamefont {Kushnir}\ \emph {et~al.}(2018)\citenamefont
  {Kushnir}, \citenamefont {Sidorenko}, \citenamefont {Tagirov},\ and\
  \citenamefont {Kupriyanov}}]{Kushnir18}%
  \BibitemOpen
  \bibfield  {author} {\bibinfo {author} {\bibfnamefont {V.~N.}\ \bibnamefont
  {Kushnir}}, \bibinfo {author} {\bibfnamefont {A.}~\bibnamefont {Sidorenko}},
  \bibinfo {author} {\bibfnamefont {L.~R.}\ \bibnamefont {Tagirov}},\ and\
  \bibinfo {author} {\bibfnamefont {M.~Y.}\ \bibnamefont {Kupriyanov}},\
  }\bibfield  {title} {\bibinfo {title} {Basic superconducting spin valves},\
  }in\ \href {https://doi.org/10.1007/978-3-319-90481-8_1} {\emph {\bibinfo
  {booktitle} {{NanoScience} and Technology}}}\ (\bibinfo  {publisher}
  {Springer International Publishing},\ \bibinfo {year} {2018})\ pp.\ \bibinfo
  {pages} {1--29}\BibitemShut {NoStop}%
\bibitem [{\citenamefont {Lenk}\ \emph {et~al.}(2017)\citenamefont {Lenk},
  \citenamefont {Morari}, \citenamefont {Zdravkov}, \citenamefont {Ullrich},
  \citenamefont {Khaydukov}, \citenamefont {Obermeier}, \citenamefont
  {M\"uller}, \citenamefont {Sidorenko}, \citenamefont {von Nidda},
  \citenamefont {Horn}, \citenamefont {Tagirov},\ and\ \citenamefont
  {Tidecks}}]{LenkPRB17}%
  \BibitemOpen
  \bibfield  {author} {\bibinfo {author} {\bibfnamefont {D.}~\bibnamefont
  {Lenk}}, \bibinfo {author} {\bibfnamefont {R.}~\bibnamefont {Morari}},
  \bibinfo {author} {\bibfnamefont {V.~I.}\ \bibnamefont {Zdravkov}}, \bibinfo
  {author} {\bibfnamefont {A.}~\bibnamefont {Ullrich}}, \bibinfo {author}
  {\bibfnamefont {Y.}~\bibnamefont {Khaydukov}}, \bibinfo {author}
  {\bibfnamefont {G.}~\bibnamefont {Obermeier}}, \bibinfo {author}
  {\bibfnamefont {C.}~\bibnamefont {M\"uller}}, \bibinfo {author}
  {\bibfnamefont {A.~S.}\ \bibnamefont {Sidorenko}}, \bibinfo {author}
  {\bibfnamefont {H.-A.~K.}\ \bibnamefont {von Nidda}}, \bibinfo {author}
  {\bibfnamefont {S.}~\bibnamefont {Horn}}, \bibinfo {author} {\bibfnamefont
  {L.~R.}\ \bibnamefont {Tagirov}},\ and\ \bibinfo {author} {\bibfnamefont
  {R.}~\bibnamefont {Tidecks}},\ }\bibfield  {title} {\bibinfo {title}
  {Full-switching {FSF}-type superconducting spin-triplet magnetic random
  access memory element},\ }\bibfield  {journal} {\bibinfo  {journal} {Physical
  Review B}\ }\textbf {\bibinfo {volume} {96}},\ \href
  {https://doi.org/10.1103/physrevb.96.184521} {10.1103/physrevb.96.184521}
  (\bibinfo {year} {2017})\BibitemShut {NoStop}%
\bibitem [{\citenamefont {Leksin}\ \emph {et~al.}(2016)\citenamefont {Leksin},
  \citenamefont {Kamashev}, \citenamefont {Schumann}, \citenamefont {Kataev},
  \citenamefont {Thomas}, \citenamefont {B\"uchner},\ and\ \citenamefont
  {Garifullin}}]{LeksinNanoRes16}%
  \BibitemOpen
  \bibfield  {author} {\bibinfo {author} {\bibfnamefont {P.~V.}\ \bibnamefont
  {Leksin}}, \bibinfo {author} {\bibfnamefont {A.~A.}\ \bibnamefont
  {Kamashev}}, \bibinfo {author} {\bibfnamefont {J.}~\bibnamefont {Schumann}},
  \bibinfo {author} {\bibfnamefont {V.~E.}\ \bibnamefont {Kataev}}, \bibinfo
  {author} {\bibfnamefont {J.}~\bibnamefont {Thomas}}, \bibinfo {author}
  {\bibfnamefont {B.}~\bibnamefont {B\"uchner}},\ and\ \bibinfo {author}
  {\bibfnamefont {I.~A.}\ \bibnamefont {Garifullin}},\ }\bibfield  {title}
  {\bibinfo {title} {Boosting the superconducting spin valve effect in a
  metallic superconductor/ferromagnet heterostructure},\ }\href
  {https://doi.org/10.1007/s12274-016-0988-y} {\bibfield  {journal} {\bibinfo
  {journal} {Nano Research}\ }\textbf {\bibinfo {volume} {9}},\ \bibinfo
  {pages} {1005} (\bibinfo {year} {2016})}\BibitemShut {NoStop}%
\bibitem [{\citenamefont {Lenk}\ \emph {et~al.}(2016)\citenamefont {Lenk},
  \citenamefont {Zdravkov}, \citenamefont {Kehrle}, \citenamefont {Obermeier},
  \citenamefont {Ullrich}, \citenamefont {Morari}, \citenamefont {von Nidda},
  \citenamefont {M\"uller}, \citenamefont {Kupriyanov}, \citenamefont
  {Sidorenko}, \citenamefont {Horn}, \citenamefont {Deminov}, \citenamefont
  {Tagirov},\ and\ \citenamefont {Tidecks}}]{LenkBJN16}%
  \BibitemOpen
  \bibfield  {author} {\bibinfo {author} {\bibfnamefont {D.}~\bibnamefont
  {Lenk}}, \bibinfo {author} {\bibfnamefont {V.~I.}\ \bibnamefont {Zdravkov}},
  \bibinfo {author} {\bibfnamefont {J.-M.}\ \bibnamefont {Kehrle}}, \bibinfo
  {author} {\bibfnamefont {G.}~\bibnamefont {Obermeier}}, \bibinfo {author}
  {\bibfnamefont {A.}~\bibnamefont {Ullrich}}, \bibinfo {author} {\bibfnamefont
  {R.}~\bibnamefont {Morari}}, \bibinfo {author} {\bibfnamefont {H.-A.~K.}\
  \bibnamefont {von Nidda}}, \bibinfo {author} {\bibfnamefont {C.}~\bibnamefont
  {M\"uller}}, \bibinfo {author} {\bibfnamefont {M.~Y.}\ \bibnamefont
  {Kupriyanov}}, \bibinfo {author} {\bibfnamefont {A.~S.}\ \bibnamefont
  {Sidorenko}}, \bibinfo {author} {\bibfnamefont {S.}~\bibnamefont {Horn}},
  \bibinfo {author} {\bibfnamefont {R.~G.}\ \bibnamefont {Deminov}}, \bibinfo
  {author} {\bibfnamefont {L.~R.}\ \bibnamefont {Tagirov}},\ and\ \bibinfo
  {author} {\bibfnamefont {R.}~\bibnamefont {Tidecks}},\ }\bibfield  {title}
  {\bibinfo {title} {Thickness dependence of the triplet spin-valve effect in
  superconductor{\textendash}ferromagnet{\textendash}ferromagnet
  heterostructures},\ }\href {https://doi.org/10.3762/bjnano.7.88} {\bibfield
  {journal} {\bibinfo  {journal} {Beilstein Journal of Nanotechnology}\
  }\textbf {\bibinfo {volume} {7}},\ \bibinfo {pages} {957} (\bibinfo {year}
  {2016})}\BibitemShut {NoStop}%
\bibitem [{\citenamefont {Stamopoulos}\ \emph {et~al.}(2014)\citenamefont
  {Stamopoulos}, \citenamefont {Aristomenopoulou},\ and\ \citenamefont
  {Manios}}]{StamopoulosAPL14}%
  \BibitemOpen
  \bibfield  {author} {\bibinfo {author} {\bibfnamefont {D.}~\bibnamefont
  {Stamopoulos}}, \bibinfo {author} {\bibfnamefont {E.}~\bibnamefont
  {Aristomenopoulou}},\ and\ \bibinfo {author} {\bibfnamefont {E.}~\bibnamefont
  {Manios}},\ }\bibfield  {title} {\bibinfo {title} {Absolute supercurrent
  switch in ferromagnetic/superconducting/ferromagnetic trilayers operating at
  t>4.2k},\ }\href@noop {} {\bibfield  {journal} {\bibinfo  {journal} {Applied
  Physics Letters}\ }\textbf {\bibinfo {volume} {105}},\ \bibinfo {pages}
  {112602} (\bibinfo {year} {2014})}\BibitemShut {NoStop}%
\bibitem [{\citenamefont {Banerjee}\ \emph {et~al.}(2014)\citenamefont
  {Banerjee}, \citenamefont {Smiet}, \citenamefont {Smits}, \citenamefont
  {Ozaeta}, \citenamefont {Bergeret}, \citenamefont {Blamire},\ and\
  \citenamefont {Robinson}}]{BanerjeeNatCom14}%
  \BibitemOpen
  \bibfield  {author} {\bibinfo {author} {\bibfnamefont {N.}~\bibnamefont
  {Banerjee}}, \bibinfo {author} {\bibfnamefont {C.~B.}\ \bibnamefont {Smiet}},
  \bibinfo {author} {\bibfnamefont {R.~G.~J.}\ \bibnamefont {Smits}}, \bibinfo
  {author} {\bibfnamefont {A.}~\bibnamefont {Ozaeta}}, \bibinfo {author}
  {\bibfnamefont {F.~S.}\ \bibnamefont {Bergeret}}, \bibinfo {author}
  {\bibfnamefont {M.~G.}\ \bibnamefont {Blamire}},\ and\ \bibinfo {author}
  {\bibfnamefont {J.~W.~A.}\ \bibnamefont {Robinson}},\ }\bibfield  {title}
  {\bibinfo {title} {Evidence for spin selectivity of triplet pairs in
  superconducting spin valves},\ }\bibfield  {journal} {\bibinfo  {journal}
  {Nature Communications}\ }\textbf {\bibinfo {volume} {5}},\ \href
  {https://doi.org/10.1038/ncomms4048} {10.1038/ncomms4048} (\bibinfo {year}
  {2014})\BibitemShut {NoStop}%
\bibitem [{\citenamefont {Flokstra}\ \emph {et~al.}(2015)\citenamefont
  {Flokstra}, \citenamefont {Cunningham}, \citenamefont {Kim}, \citenamefont
  {Satchell}, \citenamefont {Burnell}, \citenamefont {Curran}, \citenamefont
  {Bending}, \citenamefont {Kinane}, \citenamefont {Cooper}, \citenamefont
  {Langridge}, \citenamefont {Isidori}, \citenamefont {Pugach}, \citenamefont
  {Eschrig},\ and\ \citenamefont {Lee}}]{FlokstraPRB15}%
  \BibitemOpen
  \bibfield  {author} {\bibinfo {author} {\bibfnamefont {M.~G.}\ \bibnamefont
  {Flokstra}}, \bibinfo {author} {\bibfnamefont {T.~C.}\ \bibnamefont
  {Cunningham}}, \bibinfo {author} {\bibfnamefont {J.}~\bibnamefont {Kim}},
  \bibinfo {author} {\bibfnamefont {N.}~\bibnamefont {Satchell}}, \bibinfo
  {author} {\bibfnamefont {G.}~\bibnamefont {Burnell}}, \bibinfo {author}
  {\bibfnamefont {P.~J.}\ \bibnamefont {Curran}}, \bibinfo {author}
  {\bibfnamefont {S.~J.}\ \bibnamefont {Bending}}, \bibinfo {author}
  {\bibfnamefont {C.~J.}\ \bibnamefont {Kinane}}, \bibinfo {author}
  {\bibfnamefont {J.~F.~K.}\ \bibnamefont {Cooper}}, \bibinfo {author}
  {\bibfnamefont {S.}~\bibnamefont {Langridge}}, \bibinfo {author}
  {\bibfnamefont {A.}~\bibnamefont {Isidori}}, \bibinfo {author} {\bibfnamefont
  {N.}~\bibnamefont {Pugach}}, \bibinfo {author} {\bibfnamefont
  {M.}~\bibnamefont {Eschrig}},\ and\ \bibinfo {author} {\bibfnamefont {S.~L.}\
  \bibnamefont {Lee}},\ }\bibfield  {title} {\bibinfo {title} {Controlled
  suppression of superconductivity by the generation of polarized cooper pairs
  in spin-valve structures},\ }\bibfield  {journal} {\bibinfo  {journal}
  {Physical Review B}\ }\textbf {\bibinfo {volume} {91}},\ \href
  {https://doi.org/10.1103/physrevb.91.060501} {10.1103/physrevb.91.060501}
  (\bibinfo {year} {2015})\BibitemShut {NoStop}%
\bibitem [{\citenamefont {Zdravkov}\ \emph
  {et~al.}(2013{\natexlab{a}})\citenamefont {Zdravkov}, \citenamefont {Lenk},
  \citenamefont {Morari}, \citenamefont {Ullrich}, \citenamefont {Obermeier},
  \citenamefont {M\"uller}, \citenamefont {von Nidda}, \citenamefont
  {Sidorenko}, \citenamefont {Horn}, \citenamefont {Tidecks},\ and\
  \citenamefont {Tagirov}}]{ZdravkovAPL13}%
  \BibitemOpen
  \bibfield  {author} {\bibinfo {author} {\bibfnamefont {V.~I.}\ \bibnamefont
  {Zdravkov}}, \bibinfo {author} {\bibfnamefont {D.}~\bibnamefont {Lenk}},
  \bibinfo {author} {\bibfnamefont {R.}~\bibnamefont {Morari}}, \bibinfo
  {author} {\bibfnamefont {A.}~\bibnamefont {Ullrich}}, \bibinfo {author}
  {\bibfnamefont {G.}~\bibnamefont {Obermeier}}, \bibinfo {author}
  {\bibfnamefont {C.}~\bibnamefont {M\"uller}}, \bibinfo {author}
  {\bibfnamefont {H.-A.~K.}\ \bibnamefont {von Nidda}}, \bibinfo {author}
  {\bibfnamefont {A.~S.}\ \bibnamefont {Sidorenko}}, \bibinfo {author}
  {\bibfnamefont {S.}~\bibnamefont {Horn}}, \bibinfo {author} {\bibfnamefont
  {R.}~\bibnamefont {Tidecks}},\ and\ \bibinfo {author} {\bibfnamefont {L.~R.}\
  \bibnamefont {Tagirov}},\ }\bibfield  {title} {\bibinfo {title} {Memory
  effect and triplet pairing generation in the superconducting exchange biased
  co/{CoOx}/cu41ni59/nb/cu41ni59layered heterostructure},\ }\href
  {https://doi.org/10.1063/1.4818266} {\bibfield  {journal} {\bibinfo
  {journal} {Applied Physics Letters}\ }\textbf {\bibinfo {volume} {103}},\
  \bibinfo {pages} {062604} (\bibinfo {year} {2013}{\natexlab{a}})}\BibitemShut
  {NoStop}%
\bibitem [{\citenamefont {Zdravkov}\ \emph
  {et~al.}(2013{\natexlab{b}})\citenamefont {Zdravkov}, \citenamefont {Kehrle},
  \citenamefont {Obermeier}, \citenamefont {Lenk}, \citenamefont {von Nidda},
  \citenamefont {M?ller}, \citenamefont {Kupriyanov}, \citenamefont
  {Sidorenko}, \citenamefont {Horn}, \citenamefont {Tidecks},\ and\
  \citenamefont {Tagirov}}]{ZdravkovPRB13}%
  \BibitemOpen
  \bibfield  {author} {\bibinfo {author} {\bibfnamefont {V.~I.}\ \bibnamefont
  {Zdravkov}}, \bibinfo {author} {\bibfnamefont {J.}~\bibnamefont {Kehrle}},
  \bibinfo {author} {\bibfnamefont {G.}~\bibnamefont {Obermeier}}, \bibinfo
  {author} {\bibfnamefont {D.}~\bibnamefont {Lenk}}, \bibinfo {author}
  {\bibfnamefont {H.-A.~K.}\ \bibnamefont {von Nidda}}, \bibinfo {author}
  {\bibfnamefont {C.}~\bibnamefont {M?ller}}, \bibinfo {author} {\bibfnamefont
  {M.~Y.}\ \bibnamefont {Kupriyanov}}, \bibinfo {author} {\bibfnamefont
  {A.~S.}\ \bibnamefont {Sidorenko}}, \bibinfo {author} {\bibfnamefont
  {S.}~\bibnamefont {Horn}}, \bibinfo {author} {\bibfnamefont {R.}~\bibnamefont
  {Tidecks}},\ and\ \bibinfo {author} {\bibfnamefont {L.~R.}\ \bibnamefont
  {Tagirov}},\ }\bibfield  {title} {\bibinfo {title} {Experimental observation
  of the triplet spin-valve effect in a superconductor-ferromagnet
  heterostructure},\ }\bibfield  {journal} {\bibinfo  {journal} {Physical
  Review B}\ }\textbf {\bibinfo {volume} {87}},\ \href
  {https://doi.org/10.1103/physrevb.87.144507} {10.1103/physrevb.87.144507}
  (\bibinfo {year} {2013}{\natexlab{b}})\BibitemShut {NoStop}%
\bibitem [{\citenamefont {Leksin}\ \emph
  {et~al.}(2012{\natexlab{a}})\citenamefont {Leksin}, \citenamefont
  {Garif'yanov}, \citenamefont {Garifullin}, \citenamefont {Schumann},
  \citenamefont {Kataev}, \citenamefont {Schmidt},\ and\ \citenamefont
  {B\"uchner}}]{LeksinPRB12}%
  \BibitemOpen
  \bibfield  {author} {\bibinfo {author} {\bibfnamefont {P.~V.}\ \bibnamefont
  {Leksin}}, \bibinfo {author} {\bibfnamefont {N.~N.}\ \bibnamefont
  {Garif'yanov}}, \bibinfo {author} {\bibfnamefont {I.~A.}\ \bibnamefont
  {Garifullin}}, \bibinfo {author} {\bibfnamefont {J.}~\bibnamefont
  {Schumann}}, \bibinfo {author} {\bibfnamefont {V.}~\bibnamefont {Kataev}},
  \bibinfo {author} {\bibfnamefont {O.~G.}\ \bibnamefont {Schmidt}},\ and\
  \bibinfo {author} {\bibfnamefont {B.}~\bibnamefont {B\"uchner}},\ }\bibfield
  {title} {\bibinfo {title} {Physical properties of the superconducting
  spin-valve fe/cu/fe/in heterostructure},\ }\bibfield  {journal} {\bibinfo
  {journal} {Physical Review B}\ }\textbf {\bibinfo {volume} {85}},\ \href
  {https://doi.org/10.1103/physrevb.85.024502} {10.1103/physrevb.85.024502}
  (\bibinfo {year} {2012}{\natexlab{a}})\BibitemShut {NoStop}%
\bibitem [{\citenamefont {Leksin}\ \emph
  {et~al.}(2012{\natexlab{b}})\citenamefont {Leksin}, \citenamefont
  {Garif'yanov}, \citenamefont {Garifullin}, \citenamefont {Fominov},
  \citenamefont {Schumann}, \citenamefont {Krupskaya}, \citenamefont {Kataev},
  \citenamefont {Schmidt},\ and\ \citenamefont {B\"uchner}}]{LeksinPRL12}%
  \BibitemOpen
  \bibfield  {author} {\bibinfo {author} {\bibfnamefont {P.~V.}\ \bibnamefont
  {Leksin}}, \bibinfo {author} {\bibfnamefont {N.~N.}\ \bibnamefont
  {Garif'yanov}}, \bibinfo {author} {\bibfnamefont {I.~A.}\ \bibnamefont
  {Garifullin}}, \bibinfo {author} {\bibfnamefont {Y.~V.}\ \bibnamefont
  {Fominov}}, \bibinfo {author} {\bibfnamefont {J.}~\bibnamefont {Schumann}},
  \bibinfo {author} {\bibfnamefont {Y.}~\bibnamefont {Krupskaya}}, \bibinfo
  {author} {\bibfnamefont {V.}~\bibnamefont {Kataev}}, \bibinfo {author}
  {\bibfnamefont {O.~G.}\ \bibnamefont {Schmidt}},\ and\ \bibinfo {author}
  {\bibfnamefont {B.}~\bibnamefont {B\"uchner}},\ }\bibfield  {title} {\bibinfo
  {title} {Evidence for triplet superconductivity in a
  superconductor-ferromagnet spin valve},\ }\bibfield  {journal} {\bibinfo
  {journal} {Physical Review Letters}\ }\textbf {\bibinfo {volume} {109}},\
  \href {https://doi.org/10.1103/physrevlett.109.057005}
  {10.1103/physrevlett.109.057005} (\bibinfo {year}
  {2012}{\natexlab{b}})\BibitemShut {NoStop}%
\bibitem [{\citenamefont {Fitzsimmons}\ \emph {et~al.}(2000)\citenamefont
  {Fitzsimmons}, \citenamefont {Yashar}, \citenamefont {Leighton},
  \citenamefont {Schuller}, \citenamefont {Nogu\'es}, \citenamefont
  {Majkrzak},\ and\ \citenamefont {Dura}}]{Fitzsimmons2000}%
  \BibitemOpen
  \bibfield  {author} {\bibinfo {author} {\bibfnamefont {M.~R.}\ \bibnamefont
  {Fitzsimmons}}, \bibinfo {author} {\bibfnamefont {P.}~\bibnamefont {Yashar}},
  \bibinfo {author} {\bibfnamefont {C.}~\bibnamefont {Leighton}}, \bibinfo
  {author} {\bibfnamefont {I.~K.}\ \bibnamefont {Schuller}}, \bibinfo {author}
  {\bibfnamefont {J.}~\bibnamefont {Nogu\'es}}, \bibinfo {author}
  {\bibfnamefont {C.~F.}\ \bibnamefont {Majkrzak}},\ and\ \bibinfo {author}
  {\bibfnamefont {J.~A.}\ \bibnamefont {Dura}},\ }\bibfield  {title} {\bibinfo
  {title} {Asymmetric magnetization reversal in exchange-biased hysteresis
  loops},\ }\href {https://doi.org/10.1103/PhysRevLett.84.3986} {\bibfield
  {journal} {\bibinfo  {journal} {Phys. Rev. Lett.}\ }\textbf {\bibinfo
  {volume} {84}},\ \bibinfo {pages} {3986} (\bibinfo {year}
  {2000})}\BibitemShut {NoStop}%
\bibitem [{\citenamefont {Leighton}\ \emph {et~al.}(2001)\citenamefont
  {Leighton}, \citenamefont {Fitzsimmons}, \citenamefont {Yashar},
  \citenamefont {Hoffmann}, \citenamefont {Nogu\'es}, \citenamefont {Dura},
  \citenamefont {Majkrzak},\ and\ \citenamefont {Schuller}}]{LeightonPRL01}%
  \BibitemOpen
  \bibfield  {author} {\bibinfo {author} {\bibfnamefont {C.}~\bibnamefont
  {Leighton}}, \bibinfo {author} {\bibfnamefont {M.~R.}\ \bibnamefont
  {Fitzsimmons}}, \bibinfo {author} {\bibfnamefont {P.}~\bibnamefont {Yashar}},
  \bibinfo {author} {\bibfnamefont {A.}~\bibnamefont {Hoffmann}}, \bibinfo
  {author} {\bibfnamefont {J.}~\bibnamefont {Nogu\'es}}, \bibinfo {author}
  {\bibfnamefont {J.}~\bibnamefont {Dura}}, \bibinfo {author} {\bibfnamefont
  {C.~F.}\ \bibnamefont {Majkrzak}},\ and\ \bibinfo {author} {\bibfnamefont
  {I.~K.}\ \bibnamefont {Schuller}},\ }\bibfield  {title} {\bibinfo {title}
  {Two-stage magnetization reversal in exchange biased bilayers},\ }\href
  {https://doi.org/10.1103/PhysRevLett.86.4394} {\bibfield  {journal} {\bibinfo
   {journal} {Phys. Rev. Lett.}\ }\textbf {\bibinfo {volume} {86}},\ \bibinfo
  {pages} {4394} (\bibinfo {year} {2001})}\BibitemShut {NoStop}%
\bibitem [{\citenamefont {Lee}\ \emph {et~al.}(2002)\citenamefont {Lee},
  \citenamefont {te~Velthuis}, \citenamefont {Felcher}, \citenamefont {Klose},
  \citenamefont {Gredig},\ and\ \citenamefont {Dahlberg}}]{LeePRB02}%
  \BibitemOpen
  \bibfield  {author} {\bibinfo {author} {\bibfnamefont {W.-T.}\ \bibnamefont
  {Lee}}, \bibinfo {author} {\bibfnamefont {S.~G.~E.}\ \bibnamefont
  {te~Velthuis}}, \bibinfo {author} {\bibfnamefont {G.~P.}\ \bibnamefont
  {Felcher}}, \bibinfo {author} {\bibfnamefont {F.}~\bibnamefont {Klose}},
  \bibinfo {author} {\bibfnamefont {T.}~\bibnamefont {Gredig}},\ and\ \bibinfo
  {author} {\bibfnamefont {E.~D.}\ \bibnamefont {Dahlberg}},\ }\bibfield
  {title} {\bibinfo {title} {Ferromagnetic domain distribution in thin films
  during magnetization reversal},\ }\href
  {https://doi.org/10.1103/PhysRevB.65.224417} {\bibfield  {journal} {\bibinfo
  {journal} {Phys. Rev. B}\ }\textbf {\bibinfo {volume} {65}},\ \bibinfo
  {pages} {224417} (\bibinfo {year} {2002})}\BibitemShut {NoStop}%
\bibitem [{\citenamefont {Radu}\ \emph {et~al.}(2002)\citenamefont {Radu},
  \citenamefont {Etzkorn}, \citenamefont {Leiner}, \citenamefont {Schmitte},
  \citenamefont {Schreyer}, \citenamefont {Westerholt},\ and\ \citenamefont
  {Zabel}}]{RaduApplPhys02}%
  \BibitemOpen
  \bibfield  {author} {\bibinfo {author} {\bibfnamefont {F.}~\bibnamefont
  {Radu}}, \bibinfo {author} {\bibfnamefont {M.}~\bibnamefont {Etzkorn}},
  \bibinfo {author} {\bibfnamefont {V.}~\bibnamefont {Leiner}}, \bibinfo
  {author} {\bibfnamefont {T.}~\bibnamefont {Schmitte}}, \bibinfo {author}
  {\bibfnamefont {A.}~\bibnamefont {Schreyer}}, \bibinfo {author}
  {\bibfnamefont {K.}~\bibnamefont {Westerholt}},\ and\ \bibinfo {author}
  {\bibfnamefont {H.}~\bibnamefont {Zabel}},\ }\bibfield  {title} {\bibinfo
  {title} {Polarised neutron reflectometry study of co/coo exchange-biased
  multilayers},\ }\href {https://doi.org/10.1007/s003390201709} {\bibfield
  {journal} {\bibinfo  {journal} {Applied Physics A}\ }\textbf {\bibinfo
  {volume} {74}},\ \bibinfo {pages} {s1570} (\bibinfo {year}
  {2002})}\BibitemShut {NoStop}%
\bibitem [{\citenamefont {Radu}\ \emph
  {et~al.}(2003{\natexlab{a}})\citenamefont {Radu}, \citenamefont {Vorobiev},
  \citenamefont {Major}, \citenamefont {Humblot}, \citenamefont {Westerholt},\
  and\ \citenamefont {Zabel}}]{RaduPhysB03}%
  \BibitemOpen
  \bibfield  {author} {\bibinfo {author} {\bibfnamefont {F.}~\bibnamefont
  {Radu}}, \bibinfo {author} {\bibfnamefont {A.}~\bibnamefont {Vorobiev}},
  \bibinfo {author} {\bibfnamefont {J.}~\bibnamefont {Major}}, \bibinfo
  {author} {\bibfnamefont {H.}~\bibnamefont {Humblot}}, \bibinfo {author}
  {\bibfnamefont {K.}~\bibnamefont {Westerholt}},\ and\ \bibinfo {author}
  {\bibfnamefont {H.}~\bibnamefont {Zabel}},\ }\bibfield  {title} {\bibinfo
  {title} {Spin-resolved off-specular neutron scattering from magnetic domain
  walls using the polarized 3he gas spin filter},\ }\href@noop {} {\bibfield
  {journal} {\bibinfo  {journal} {Physica B: Condensed Matter}\ }\textbf
  {\bibinfo {volume} {335}},\ \bibinfo {pages} {63} (\bibinfo {year}
  {2003}{\natexlab{a}})}\BibitemShut {NoStop}%
\bibitem [{\citenamefont {Radu}\ \emph
  {et~al.}(2003{\natexlab{b}})\citenamefont {Radu}, \citenamefont {Etzkorn},
  \citenamefont {Siebrecht}, \citenamefont {Schmitte}, \citenamefont
  {Westerholt},\ and\ \citenamefont {Zabel}}]{RaduPRB03}%
  \BibitemOpen
  \bibfield  {author} {\bibinfo {author} {\bibfnamefont {F.}~\bibnamefont
  {Radu}}, \bibinfo {author} {\bibfnamefont {M.}~\bibnamefont {Etzkorn}},
  \bibinfo {author} {\bibfnamefont {R.}~\bibnamefont {Siebrecht}}, \bibinfo
  {author} {\bibfnamefont {T.}~\bibnamefont {Schmitte}}, \bibinfo {author}
  {\bibfnamefont {K.}~\bibnamefont {Westerholt}},\ and\ \bibinfo {author}
  {\bibfnamefont {H.}~\bibnamefont {Zabel}},\ }\bibfield  {title} {\bibinfo
  {title} {Interfacial domain formation during magnetization reversal in
  exchange-biased coo/co bilayers},\ }\href
  {https://doi.org/10.1103/PhysRevB.67.134409} {\bibfield  {journal} {\bibinfo
  {journal} {Phys. Rev. B}\ }\textbf {\bibinfo {volume} {67}},\ \bibinfo
  {pages} {134409} (\bibinfo {year} {2003}{\natexlab{b}})}\BibitemShut
  {NoStop}%
\bibitem [{\citenamefont {Radu}\ \emph {et~al.}(2005)\citenamefont {Radu},
  \citenamefont {Westphalen}, \citenamefont {Theis-Br\"ohl},\ and\
  \citenamefont {Zabel}}]{RaduJOP05}%
  \BibitemOpen
  \bibfield  {author} {\bibinfo {author} {\bibfnamefont {F.}~\bibnamefont
  {Radu}}, \bibinfo {author} {\bibfnamefont {A.}~\bibnamefont {Westphalen}},
  \bibinfo {author} {\bibfnamefont {K.}~\bibnamefont {Theis-Br\"ohl}},\ and\
  \bibinfo {author} {\bibfnamefont {H.}~\bibnamefont {Zabel}},\ }\bibfield
  {title} {\bibinfo {title} {Quantitative description of the azimuthal
  dependence of the exchange bias effect},\ }\href
  {https://doi.org/10.1088/0953-8984/18/3/l01} {\bibfield  {journal} {\bibinfo
  {journal} {Journal of Physics: Condensed Matter}\ }\textbf {\bibinfo {volume}
  {18}},\ \bibinfo {pages} {L29} (\bibinfo {year} {2005})}\BibitemShut
  {NoStop}%
\bibitem [{\citenamefont {Paul}\ \emph
  {et~al.}(2006{\natexlab{a}})\citenamefont {Paul}, \citenamefont {Kentzinger},
  \citenamefont {R\"ucker},\ and\ \citenamefont {Br\"uckel}}]{PaulJOP06}%
  \BibitemOpen
  \bibfield  {author} {\bibinfo {author} {\bibfnamefont {A.}~\bibnamefont
  {Paul}}, \bibinfo {author} {\bibfnamefont {E.}~\bibnamefont {Kentzinger}},
  \bibinfo {author} {\bibfnamefont {U.}~\bibnamefont {R\"ucker}},\ and\
  \bibinfo {author} {\bibfnamefont {T.}~\bibnamefont {Br\"uckel}},\ }\bibfield
  {title} {\bibinfo {title} {The angular dependence of the magnetization
  reversal in exchange biased multilayers},\ }\href
  {https://doi.org/10.1088/0953-8984/18/12/l01} {\bibfield  {journal} {\bibinfo
   {journal} {Journal of Physics: Condensed Matter}\ }\textbf {\bibinfo
  {volume} {18}},\ \bibinfo {pages} {L149} (\bibinfo {year}
  {2006}{\natexlab{a}})}\BibitemShut {NoStop}%
\bibitem [{\citenamefont {Paul}\ \emph
  {et~al.}(2006{\natexlab{b}})\citenamefont {Paul}, \citenamefont {Kentzinger},
  \citenamefont {R\"ucker},\ and\ \citenamefont {Br\"uckel}}]{PaulPRB06}%
  \BibitemOpen
  \bibfield  {author} {\bibinfo {author} {\bibfnamefont {A.}~\bibnamefont
  {Paul}}, \bibinfo {author} {\bibfnamefont {E.}~\bibnamefont {Kentzinger}},
  \bibinfo {author} {\bibfnamefont {U.}~\bibnamefont {R\"ucker}},\ and\
  \bibinfo {author} {\bibfnamefont {T.}~\bibnamefont {Br\"uckel}},\ }\bibfield
  {title} {\bibinfo {title} {Magnetization reversal with variation of the ratio
  of the anisotropy energies in exchange bias systems},\ }\href
  {https://doi.org/10.1103/PhysRevB.74.054424} {\bibfield  {journal} {\bibinfo
  {journal} {Phys. Rev. B}\ }\textbf {\bibinfo {volume} {74}},\ \bibinfo
  {pages} {054424} (\bibinfo {year} {2006}{\natexlab{b}})}\BibitemShut
  {NoStop}%
\bibitem [{\citenamefont {Paul}\ \emph {et~al.}(2007)\citenamefont {Paul},
  \citenamefont {Br\"uckel}, \citenamefont {Kentzinger},\ and\ \citenamefont
  {R\"ucker}}]{PaulJOP07}%
  \BibitemOpen
  \bibfield  {author} {\bibinfo {author} {\bibfnamefont {A.}~\bibnamefont
  {Paul}}, \bibinfo {author} {\bibfnamefont {T.}~\bibnamefont {Br\"uckel}},
  \bibinfo {author} {\bibfnamefont {E.}~\bibnamefont {Kentzinger}},\ and\
  \bibinfo {author} {\bibfnamefont {U.}~\bibnamefont {R\"ucker}},\ }\bibfield
  {title} {\bibinfo {title} {Magnetization reversal in trained exchange biased
  multilayers},\ }\href {https://doi.org/10.1088/0953-8984/19/8/086229}
  {\bibfield  {journal} {\bibinfo  {journal} {Journal of Physics: Condensed
  Matter}\ }\textbf {\bibinfo {volume} {19}},\ \bibinfo {pages} {086229}
  (\bibinfo {year} {2007})}\BibitemShut {NoStop}%
\bibitem [{\citenamefont {Paul}\ and\ \citenamefont
  {Mattauch}(2010)}]{PaulJAP10}%
  \BibitemOpen
  \bibfield  {author} {\bibinfo {author} {\bibfnamefont {A.}~\bibnamefont
  {Paul}}\ and\ \bibinfo {author} {\bibfnamefont {S.}~\bibnamefont
  {Mattauch}},\ }\bibfield  {title} {\bibinfo {title} {Can uniaxial anisotropy
  be responsible for training in exchange coupled system?},\ }\href@noop {}
  {\bibfield  {journal} {\bibinfo  {journal} {Journal of applied physics}\
  }\textbf {\bibinfo {volume} {108}},\ \bibinfo {pages} {053918} (\bibinfo
  {year} {2010})}\BibitemShut {NoStop}%
\bibitem [{\citenamefont {Cortie}\ \emph {et~al.}(2012)\citenamefont {Cortie},
  \citenamefont {Lin}, \citenamefont {Shueh}, \citenamefont {Hsu},
  \citenamefont {Wang}, \citenamefont {James}, \citenamefont {Fritzsche},
  \citenamefont {Br\"uck},\ and\ \citenamefont {Klose}}]{Cortie12}%
  \BibitemOpen
  \bibfield  {author} {\bibinfo {author} {\bibfnamefont {D.~L.}\ \bibnamefont
  {Cortie}}, \bibinfo {author} {\bibfnamefont {K.-W.}\ \bibnamefont {Lin}},
  \bibinfo {author} {\bibfnamefont {C.}~\bibnamefont {Shueh}}, \bibinfo
  {author} {\bibfnamefont {H.-F.}\ \bibnamefont {Hsu}}, \bibinfo {author}
  {\bibfnamefont {X.~L.}\ \bibnamefont {Wang}}, \bibinfo {author}
  {\bibfnamefont {M.}~\bibnamefont {James}}, \bibinfo {author} {\bibfnamefont
  {H.}~\bibnamefont {Fritzsche}}, \bibinfo {author} {\bibfnamefont
  {S.}~\bibnamefont {Br\"uck}},\ and\ \bibinfo {author} {\bibfnamefont
  {F.}~\bibnamefont {Klose}},\ }\bibfield  {title} {\bibinfo {title} {Exchange
  bias in a nanocrystalline hematite/permalloy thin film investigated with
  polarized neutron reflectometry},\ }\bibfield  {journal} {\bibinfo  {journal}
  {Physical Review B}\ }\textbf {\bibinfo {volume} {86}},\ \href
  {https://doi.org/10.1103/physrevb.86.054408} {10.1103/physrevb.86.054408}
  (\bibinfo {year} {2012})\BibitemShut {NoStop}%
\bibitem [{\citenamefont {Demeter}\ \emph {et~al.}(2012)\citenamefont
  {Demeter}, \citenamefont {Men{\'{e}}ndez}, \citenamefont {Teichert},
  \citenamefont {Steitz}, \citenamefont {Paramanik}, \citenamefont
  {Haesendonck}, \citenamefont {Vantomme},\ and\ \citenamefont
  {Temst}}]{Demeter12}%
  \BibitemOpen
  \bibfield  {author} {\bibinfo {author} {\bibfnamefont {J.}~\bibnamefont
  {Demeter}}, \bibinfo {author} {\bibfnamefont {E.}~\bibnamefont
  {Men{\'{e}}ndez}}, \bibinfo {author} {\bibfnamefont {A.}~\bibnamefont
  {Teichert}}, \bibinfo {author} {\bibfnamefont {R.}~\bibnamefont {Steitz}},
  \bibinfo {author} {\bibfnamefont {D.}~\bibnamefont {Paramanik}}, \bibinfo
  {author} {\bibfnamefont {C.~V.}\ \bibnamefont {Haesendonck}}, \bibinfo
  {author} {\bibfnamefont {A.}~\bibnamefont {Vantomme}},\ and\ \bibinfo
  {author} {\bibfnamefont {K.}~\bibnamefont {Temst}},\ }\bibfield  {title}
  {\bibinfo {title} {Influence of magnetocrystalline anisotropy on the
  magnetization reversal mechanism in exchange bias co/{CoO} bilayers},\ }\href
  {https://doi.org/10.1016/j.ssc.2011.11.026} {\bibfield  {journal} {\bibinfo
  {journal} {Solid State Communications}\ }\textbf {\bibinfo {volume} {152}},\
  \bibinfo {pages} {292} (\bibinfo {year} {2012})}\BibitemShut {NoStop}%
\bibitem [{\citenamefont {Paul}\ \emph {et~al.}(2013)\citenamefont {Paul},
  \citenamefont {Paul}, \citenamefont {Jutimoosik}, \citenamefont {Yimnirun},
  \citenamefont {Rujirawat}, \citenamefont {H\"opfner}, \citenamefont
  {Lauermann}, \citenamefont {Lux-Steiner}, \citenamefont {Mattauch},\ and\
  \citenamefont {B\"oni}}]{PaulPRB13}%
  \BibitemOpen
  \bibfield  {author} {\bibinfo {author} {\bibfnamefont {A.}~\bibnamefont
  {Paul}}, \bibinfo {author} {\bibfnamefont {N.}~\bibnamefont {Paul}}, \bibinfo
  {author} {\bibfnamefont {J.}~\bibnamefont {Jutimoosik}}, \bibinfo {author}
  {\bibfnamefont {R.}~\bibnamefont {Yimnirun}}, \bibinfo {author}
  {\bibfnamefont {S.}~\bibnamefont {Rujirawat}}, \bibinfo {author}
  {\bibfnamefont {B.}~\bibnamefont {H\"opfner}}, \bibinfo {author}
  {\bibfnamefont {I.}~\bibnamefont {Lauermann}}, \bibinfo {author}
  {\bibfnamefont {M.}~\bibnamefont {Lux-Steiner}}, \bibinfo {author}
  {\bibfnamefont {S.}~\bibnamefont {Mattauch}},\ and\ \bibinfo {author}
  {\bibfnamefont {P.}~\bibnamefont {B\"oni}},\ }\bibfield  {title} {\bibinfo
  {title} {Change in interface magnetism of an exchange-coupled system due to
  the presence of nonmagnetic spacers},\ }\href
  {https://doi.org/10.1103/PhysRevB.87.014431} {\bibfield  {journal} {\bibinfo
  {journal} {Phys. Rev. B}\ }\textbf {\bibinfo {volume} {87}},\ \bibinfo
  {pages} {014431} (\bibinfo {year} {2013})}\BibitemShut {NoStop}%
\bibitem [{\citenamefont {Khaydukov}\ \emph
  {et~al.}(2017{\natexlab{a}})\citenamefont {Khaydukov}, \citenamefont
  {Morari}, \citenamefont {Zdravkov}, \citenamefont {Mustafa}, \citenamefont
  {Keller}, \citenamefont {Keimer},\ and\ \citenamefont
  {Sidorenko}}]{KhaydukovLowT17}%
  \BibitemOpen
  \bibfield  {author} {\bibinfo {author} {\bibfnamefont {Y.}~\bibnamefont
  {Khaydukov}}, \bibinfo {author} {\bibfnamefont {R.}~\bibnamefont {Morari}},
  \bibinfo {author} {\bibfnamefont {V.}~\bibnamefont {Zdravkov}}, \bibinfo
  {author} {\bibfnamefont {L.}~\bibnamefont {Mustafa}}, \bibinfo {author}
  {\bibfnamefont {T.}~\bibnamefont {Keller}}, \bibinfo {author} {\bibfnamefont
  {B.}~\bibnamefont {Keimer}},\ and\ \bibinfo {author} {\bibfnamefont
  {A.}~\bibnamefont {Sidorenko}},\ }\bibfield  {title} {\bibinfo {title}
  {Evolution of non-collinear magnetic state of exchange biased
  ferromagnet/normal metal/ferromagnet/superconductor heterostructure in
  magnetic field studied by polarized neutron reflectometry},\ }\href@noop {}
  {\bibfield  {journal} {\bibinfo  {journal} {Low Temperature Physics}\
  }\textbf {\bibinfo {volume} {43}},\ \bibinfo {pages} {837} (\bibinfo {year}
  {2017}{\natexlab{a}})}\BibitemShut {NoStop}%
\bibitem [{\citenamefont {Kim}\ \emph {et~al.}(2019)\citenamefont {Kim},
  \citenamefont {Khaydukov}, \citenamefont {Bluschke}, \citenamefont {Suyolcu},
  \citenamefont {Christiani}, \citenamefont {Son}, \citenamefont {Dietl},
  \citenamefont {Keller}, \citenamefont {Weschke}, \citenamefont {van Aken},
  \citenamefont {Logvenov},\ and\ \citenamefont {Keimer}}]{Kim19}%
  \BibitemOpen
  \bibfield  {author} {\bibinfo {author} {\bibfnamefont {G.}~\bibnamefont
  {Kim}}, \bibinfo {author} {\bibfnamefont {Y.}~\bibnamefont {Khaydukov}},
  \bibinfo {author} {\bibfnamefont {M.}~\bibnamefont {Bluschke}}, \bibinfo
  {author} {\bibfnamefont {Y.~E.}\ \bibnamefont {Suyolcu}}, \bibinfo {author}
  {\bibfnamefont {G.}~\bibnamefont {Christiani}}, \bibinfo {author}
  {\bibfnamefont {K.}~\bibnamefont {Son}}, \bibinfo {author} {\bibfnamefont
  {C.}~\bibnamefont {Dietl}}, \bibinfo {author} {\bibfnamefont
  {T.}~\bibnamefont {Keller}}, \bibinfo {author} {\bibfnamefont
  {E.}~\bibnamefont {Weschke}}, \bibinfo {author} {\bibfnamefont {P.~A.}\
  \bibnamefont {van Aken}}, \bibinfo {author} {\bibfnamefont {G.}~\bibnamefont
  {Logvenov}},\ and\ \bibinfo {author} {\bibfnamefont {B.}~\bibnamefont
  {Keimer}},\ }\bibfield  {title} {\bibinfo {title} {Tunable perpendicular
  exchange bias in oxide heterostructures},\ }\bibfield  {journal} {\bibinfo
  {journal} {Physical Review Materials}\ }\textbf {\bibinfo {volume} {3}},\
  \href {https://doi.org/10.1103/physrevmaterials.3.084420}
  {10.1103/physrevmaterials.3.084420} (\bibinfo {year} {2019})\BibitemShut
  {NoStop}%
\bibitem [{\citenamefont {Chen}\ \emph {et~al.}(2019)\citenamefont {Chen},
  \citenamefont {Philippi-Kobs}, \citenamefont {Lauter}, \citenamefont
  {Vorobiev}, \citenamefont {Dyadkina}, \citenamefont {Yakovchuk},
  \citenamefont {Stolyar},\ and\ \citenamefont {Lott}}]{Chen19}%
  \BibitemOpen
  \bibfield  {author} {\bibinfo {author} {\bibfnamefont {K.}~\bibnamefont
  {Chen}}, \bibinfo {author} {\bibfnamefont {A.}~\bibnamefont {Philippi-Kobs}},
  \bibinfo {author} {\bibfnamefont {V.}~\bibnamefont {Lauter}}, \bibinfo
  {author} {\bibfnamefont {A.}~\bibnamefont {Vorobiev}}, \bibinfo {author}
  {\bibfnamefont {E.}~\bibnamefont {Dyadkina}}, \bibinfo {author}
  {\bibfnamefont {V.}~\bibnamefont {Yakovchuk}}, \bibinfo {author}
  {\bibfnamefont {S.}~\bibnamefont {Stolyar}},\ and\ \bibinfo {author}
  {\bibfnamefont {D.}~\bibnamefont {Lott}},\ }\bibfield  {title} {\bibinfo
  {title} {Observation of a chirality-induced exchange-bias effect},\
  }\bibfield  {journal} {\bibinfo  {journal} {Physical Review Applied}\
  }\textbf {\bibinfo {volume} {12}},\ \href
  {https://doi.org/10.1103/physrevapplied.12.024047}
  {10.1103/physrevapplied.12.024047} (\bibinfo {year} {2019})\BibitemShut
  {NoStop}%
\bibitem [{\citenamefont {Zabel}\ \emph {et~al.}(2007)\citenamefont {Zabel},
  \citenamefont {Theis-Br\"ohl},\ and\ \citenamefont {Toperverg}}]{Zabel07}%
  \BibitemOpen
  \bibfield  {author} {\bibinfo {author} {\bibfnamefont {H.}~\bibnamefont
  {Zabel}}, \bibinfo {author} {\bibfnamefont {K.}~\bibnamefont
  {Theis-Br\"ohl}},\ and\ \bibinfo {author} {\bibfnamefont {B.~P.}\
  \bibnamefont {Toperverg}},\ }\bibinfo {title} {Polarized neutron reflectivity
  and scattering from magnetic nanostructures and spintronic materials},\ in\
  \href {https://doi.org/10.1002/9780470022184.hmm303} {\emph {\bibinfo
  {booktitle} {Handbook of Magnetism and Advanced Magnetic Materials}}}\
  (\bibinfo  {publisher} {American Cancer Society},\ \bibinfo {year}
  {2007})\BibitemShut {NoStop}%
\bibitem [{\citenamefont {Toperverg}(2015)}]{Toperverg15}%
  \BibitemOpen
  \bibfield  {author} {\bibinfo {author} {\bibfnamefont {B.~P.}\ \bibnamefont
  {Toperverg}},\ }\bibfield  {title} {\bibinfo {title} {Polarized neutron
  reflectometry of magnetic nanostructures},\ }\href@noop {} {\bibfield
  {journal} {\bibinfo  {journal} {The Physics of Metals and Metallography}\
  }\textbf {\bibinfo {volume} {116}},\ \bibinfo {pages} {1337} (\bibinfo {year}
  {2015})}\BibitemShut {NoStop}%
\bibitem [{\citenamefont {Khaydukov}\ \emph {et~al.}(2014)\citenamefont
  {Khaydukov}, \citenamefont {Ovsyannikov}, \citenamefont {Sheyerman},
  \citenamefont {Constantinian}, \citenamefont {Mustafa}, \citenamefont
  {Keller}, \citenamefont {Uribe-Laverde}, \citenamefont {Kislinskii},
  \citenamefont {Shadrin}, \citenamefont {Kalaboukhov}, \citenamefont
  {Keimer},\ and\ \citenamefont {Winkler}}]{KhaydukovPRB14}%
  \BibitemOpen
  \bibfield  {author} {\bibinfo {author} {\bibfnamefont {Y.~N.}\ \bibnamefont
  {Khaydukov}}, \bibinfo {author} {\bibfnamefont {G.~A.}\ \bibnamefont
  {Ovsyannikov}}, \bibinfo {author} {\bibfnamefont {A.~E.}\ \bibnamefont
  {Sheyerman}}, \bibinfo {author} {\bibfnamefont {K.~Y.}\ \bibnamefont
  {Constantinian}}, \bibinfo {author} {\bibfnamefont {L.}~\bibnamefont
  {Mustafa}}, \bibinfo {author} {\bibfnamefont {T.}~\bibnamefont {Keller}},
  \bibinfo {author} {\bibfnamefont {M.~A.}\ \bibnamefont {Uribe-Laverde}},
  \bibinfo {author} {\bibfnamefont {Y.~V.}\ \bibnamefont {Kislinskii}},
  \bibinfo {author} {\bibfnamefont {A.~V.}\ \bibnamefont {Shadrin}}, \bibinfo
  {author} {\bibfnamefont {A.}~\bibnamefont {Kalaboukhov}}, \bibinfo {author}
  {\bibfnamefont {B.}~\bibnamefont {Keimer}},\ and\ \bibinfo {author}
  {\bibfnamefont {D.}~\bibnamefont {Winkler}},\ }\bibfield  {title} {\bibinfo
  {title} {Evidence for spin-triplet superconducting correlations in
  metal-oxide heterostructures with noncollinear magnetization},\ }\href
  {https://doi.org/10.1103/PhysRevB.90.035130} {\bibfield  {journal} {\bibinfo
  {journal} {Phys. Rev. B}\ }\textbf {\bibinfo {volume} {90}},\ \bibinfo
  {pages} {035130} (\bibinfo {year} {2014})}\BibitemShut {NoStop}%
\bibitem [{\citenamefont {Khaydukov}\ \emph
  {et~al.}(2017{\natexlab{b}})\citenamefont {Khaydukov}, \citenamefont
  {Petrzhik}, \citenamefont {Borisenko}, \citenamefont {Kalabukhov},
  \citenamefont {Winkler}, \citenamefont {Keller}, \citenamefont
  {Ovsyannikov},\ and\ \citenamefont {Keimer}}]{KhaydukovPRB17}%
  \BibitemOpen
  \bibfield  {author} {\bibinfo {author} {\bibfnamefont {Y.}~\bibnamefont
  {Khaydukov}}, \bibinfo {author} {\bibfnamefont {A.~M.}\ \bibnamefont
  {Petrzhik}}, \bibinfo {author} {\bibfnamefont {I.~V.}\ \bibnamefont
  {Borisenko}}, \bibinfo {author} {\bibfnamefont {A.}~\bibnamefont
  {Kalabukhov}}, \bibinfo {author} {\bibfnamefont {D.}~\bibnamefont {Winkler}},
  \bibinfo {author} {\bibfnamefont {T.}~\bibnamefont {Keller}}, \bibinfo
  {author} {\bibfnamefont {G.~A.}\ \bibnamefont {Ovsyannikov}},\ and\ \bibinfo
  {author} {\bibfnamefont {B.}~\bibnamefont {Keimer}},\ }\bibfield  {title}
  {\bibinfo {title} {Magnetic waveguides for neutron reflectometry},\ }\href
  {https://doi.org/10.1103/PhysRevB.96.165414} {\bibfield  {journal} {\bibinfo
  {journal} {Phys. Rev. B}\ }\textbf {\bibinfo {volume} {96}},\ \bibinfo
  {pages} {165414} (\bibinfo {year} {2017}{\natexlab{b}})}\BibitemShut
  {NoStop}%
\bibitem [{\citenamefont {Khaydukov}\ \emph {et~al.}(2019)\citenamefont
  {Khaydukov}, \citenamefont {Kravtsov}, \citenamefont {Zhaketov},
  \citenamefont {Progliado}, \citenamefont {Kim}, \citenamefont {Nikitenko},
  \citenamefont {Keller}, \citenamefont {Ustinov}, \citenamefont {Aksenov},\
  and\ \citenamefont {Keimer}}]{KhaydukovPRB19}%
  \BibitemOpen
  \bibfield  {author} {\bibinfo {author} {\bibfnamefont {Y.~N.}\ \bibnamefont
  {Khaydukov}}, \bibinfo {author} {\bibfnamefont {E.~A.}\ \bibnamefont
  {Kravtsov}}, \bibinfo {author} {\bibfnamefont {V.~D.}\ \bibnamefont
  {Zhaketov}}, \bibinfo {author} {\bibfnamefont {V.~V.}\ \bibnamefont
  {Progliado}}, \bibinfo {author} {\bibfnamefont {G.}~\bibnamefont {Kim}},
  \bibinfo {author} {\bibfnamefont {Y.~V.}\ \bibnamefont {Nikitenko}}, \bibinfo
  {author} {\bibfnamefont {T.}~\bibnamefont {Keller}}, \bibinfo {author}
  {\bibfnamefont {V.~V.}\ \bibnamefont {Ustinov}}, \bibinfo {author}
  {\bibfnamefont {V.~L.}\ \bibnamefont {Aksenov}},\ and\ \bibinfo {author}
  {\bibfnamefont {B.}~\bibnamefont {Keimer}},\ }\bibfield  {title} {\bibinfo
  {title} {Magnetic proximity effect in nb/gd superlattices seen by neutron
  reflectometry},\ }\href {https://doi.org/10.1103/PhysRevB.99.140503}
  {\bibfield  {journal} {\bibinfo  {journal} {Phys. Rev. B}\ }\textbf {\bibinfo
  {volume} {99}},\ \bibinfo {pages} {140503} (\bibinfo {year}
  {2019})}\BibitemShut {NoStop}%
\bibitem [{\citenamefont {Welp}\ \emph {et~al.}(2003)\citenamefont {Welp},
  \citenamefont {te~Velthuis}, \citenamefont {Felcher}, \citenamefont
  {Gredig},\ and\ \citenamefont {Dahlberg}}]{WelpJAP03}%
  \BibitemOpen
  \bibfield  {author} {\bibinfo {author} {\bibfnamefont {U.}~\bibnamefont
  {Welp}}, \bibinfo {author} {\bibfnamefont {S.~G.~E.}\ \bibnamefont
  {te~Velthuis}}, \bibinfo {author} {\bibfnamefont {G.~P.}\ \bibnamefont
  {Felcher}}, \bibinfo {author} {\bibfnamefont {T.}~\bibnamefont {Gredig}},\
  and\ \bibinfo {author} {\bibfnamefont {E.~D.}\ \bibnamefont {Dahlberg}},\
  }\bibfield  {title} {\bibinfo {title} {Domain formation in exchange biased
  co/{CoO} bilayers},\ }\href {https://doi.org/10.1063/1.1540152} {\bibfield
  {journal} {\bibinfo  {journal} {Journal of Applied Physics}\ }\textbf
  {\bibinfo {volume} {93}},\ \bibinfo {pages} {7726} (\bibinfo {year}
  {2003})}\BibitemShut {NoStop}%
\bibitem [{\citenamefont {Brems}\ \emph {et~al.}(2008)\citenamefont {Brems},
  \citenamefont {Volodin}, \citenamefont {Haesendonck},\ and\ \citenamefont
  {Temst}}]{BremsJAP08}%
  \BibitemOpen
  \bibfield  {author} {\bibinfo {author} {\bibfnamefont {S.}~\bibnamefont
  {Brems}}, \bibinfo {author} {\bibfnamefont {A.}~\bibnamefont {Volodin}},
  \bibinfo {author} {\bibfnamefont {C.~V.}\ \bibnamefont {Haesendonck}},\ and\
  \bibinfo {author} {\bibfnamefont {K.}~\bibnamefont {Temst}},\ }\bibfield
  {title} {\bibinfo {title} {Magnetic force microscopy study of the training
  effect in polycrystalline co/{CoO} bilayers},\ }\href
  {https://doi.org/10.1063/1.2938035} {\bibfield  {journal} {\bibinfo
  {journal} {Journal of Applied Physics}\ }\textbf {\bibinfo {volume} {103}},\
  \bibinfo {pages} {113912} (\bibinfo {year} {2008})}\BibitemShut {NoStop}%
\bibitem [{\citenamefont {Schwink}\ and\ \citenamefont
  {Sch\"arpf}(1975)}]{Schwink75}%
  \BibitemOpen
  \bibfield  {author} {\bibinfo {author} {\bibfnamefont {C.}~\bibnamefont
  {Schwink}}\ and\ \bibinfo {author} {\bibfnamefont {O.}~\bibnamefont
  {Sch\"arpf}},\ }\bibfield  {title} {\bibinfo {title} {Solution of the
  pauli-equation for neutrons in varying magnetic fields and its application to
  reflection and transmission at helical magnetic structures},\ }\href
  {https://doi.org/10.1007/bf01313312} {\bibfield  {journal} {\bibinfo
  {journal} {Zeitschrift f\"ur Physik B Condensed Matter and Quanta}\ }\textbf
  {\bibinfo {volume} {21}},\ \bibinfo {pages} {305} (\bibinfo {year}
  {1975})}\BibitemShut {NoStop}%
\bibitem [{\citenamefont {Aksenov}\ \emph {et~al.}(2007)\citenamefont
  {Aksenov}, \citenamefont {Ignatovich},\ and\ \citenamefont
  {Nikitenko}}]{Aksenov07}%
  \BibitemOpen
  \bibfield  {author} {\bibinfo {author} {\bibfnamefont {V.~L.}\ \bibnamefont
  {Aksenov}}, \bibinfo {author} {\bibfnamefont {V.~K.}\ \bibnamefont
  {Ignatovich}},\ and\ \bibinfo {author} {\bibfnamefont {Y.~V.}\ \bibnamefont
  {Nikitenko}},\ }\bibfield  {title} {\bibinfo {title} {Reflection of neutrons
  from a helical system},\ }\href {https://doi.org/10.1134/s0021364006210016}
  {\bibfield  {journal} {\bibinfo  {journal} {{JETP} Letters}\ }\textbf
  {\bibinfo {volume} {84}},\ \bibinfo {pages} {473} (\bibinfo {year}
  {2007})}\BibitemShut {NoStop}%
\bibitem [{\citenamefont {Fraerman}\ and\ \citenamefont
  {Udalov}(2007)}]{Fraerman07}%
  \BibitemOpen
  \bibfield  {author} {\bibinfo {author} {\bibfnamefont {A.~A.}\ \bibnamefont
  {Fraerman}}\ and\ \bibinfo {author} {\bibfnamefont {O.~G.}\ \bibnamefont
  {Udalov}},\ }\bibfield  {title} {\bibinfo {title} {Specific features of the
  motion of neutrons in a medium with a helical magnetic structure},\ }\href
  {https://doi.org/10.1134/s1063776107010074} {\bibfield  {journal} {\bibinfo
  {journal} {Journal of Experimental and Theoretical Physics}\ }\textbf
  {\bibinfo {volume} {104}},\ \bibinfo {pages} {62} (\bibinfo {year}
  {2007})}\BibitemShut {NoStop}%
\bibitem [{\citenamefont {Grigoriev}\ \emph {et~al.}(2008)\citenamefont
  {Grigoriev}, \citenamefont {Chetverikov}, \citenamefont {Lott},\ and\
  \citenamefont {Schreyer}}]{Grigoriev08}%
  \BibitemOpen
  \bibfield  {author} {\bibinfo {author} {\bibfnamefont {S.~V.}\ \bibnamefont
  {Grigoriev}}, \bibinfo {author} {\bibfnamefont {Y.~O.}\ \bibnamefont
  {Chetverikov}}, \bibinfo {author} {\bibfnamefont {D.}~\bibnamefont {Lott}},\
  and\ \bibinfo {author} {\bibfnamefont {A.}~\bibnamefont {Schreyer}},\
  }\bibfield  {title} {\bibinfo {title} {Field induced chirality in the helix
  structure {ofDy}/{YMultilayer} films and experimental evidence for
  dzyaloshinskii-moriya interaction on the interfaces},\ }\bibfield  {journal}
  {\bibinfo  {journal} {Physical Review Letters}\ }\textbf {\bibinfo {volume}
  {100}},\ \href {https://doi.org/10.1103/physrevlett.100.197203}
  {10.1103/physrevlett.100.197203} (\bibinfo {year} {2008})\BibitemShut
  {NoStop}%
\bibitem [{\citenamefont {Tarnavich}\ \emph {et~al.}(2014)\citenamefont
  {Tarnavich}, \citenamefont {Lott}, \citenamefont {Mattauch}, \citenamefont
  {Oleshkevych}, \citenamefont {Kapaklis},\ and\ \citenamefont
  {Grigoriev}}]{Tarnavich14}%
  \BibitemOpen
  \bibfield  {author} {\bibinfo {author} {\bibfnamefont {V.~V.}\ \bibnamefont
  {Tarnavich}}, \bibinfo {author} {\bibfnamefont {D.}~\bibnamefont {Lott}},
  \bibinfo {author} {\bibfnamefont {S.}~\bibnamefont {Mattauch}}, \bibinfo
  {author} {\bibfnamefont {A.}~\bibnamefont {Oleshkevych}}, \bibinfo {author}
  {\bibfnamefont {V.}~\bibnamefont {Kapaklis}},\ and\ \bibinfo {author}
  {\bibfnamefont {S.~V.}\ \bibnamefont {Grigoriev}},\ }\bibfield  {title}
  {\bibinfo {title} {Field-induced chirality in the helix structure of ho/y
  multilayers},\ }\bibfield  {journal} {\bibinfo  {journal} {Physical Review
  B}\ }\textbf {\bibinfo {volume} {89}},\ \href
  {https://doi.org/10.1103/physrevb.89.054406} {10.1103/physrevb.89.054406}
  (\bibinfo {year} {2014})\BibitemShut {NoStop}%
\bibitem [{\citenamefont {Tarnavich}\ \emph {et~al.}(2017)\citenamefont
  {Tarnavich}, \citenamefont {Tartakovskaya}, \citenamefont {Chetverikov},
  \citenamefont {Golub}, \citenamefont {Lott}, \citenamefont {Chernenkov},
  \citenamefont {Devishvili}, \citenamefont {Ukleev}, \citenamefont {Kapaklis},
  \citenamefont {Oleshkevych}, \citenamefont {Fedorov}, \citenamefont
  {Bairamukov}, \citenamefont {Vorobiev},\ and\ \citenamefont
  {Grigoriev}}]{Tarnavich17}%
  \BibitemOpen
  \bibfield  {author} {\bibinfo {author} {\bibfnamefont {V.}~\bibnamefont
  {Tarnavich}}, \bibinfo {author} {\bibfnamefont {E.}~\bibnamefont
  {Tartakovskaya}}, \bibinfo {author} {\bibfnamefont {Y.}~\bibnamefont
  {Chetverikov}}, \bibinfo {author} {\bibfnamefont {V.}~\bibnamefont {Golub}},
  \bibinfo {author} {\bibfnamefont {D.}~\bibnamefont {Lott}}, \bibinfo {author}
  {\bibfnamefont {Y.}~\bibnamefont {Chernenkov}}, \bibinfo {author}
  {\bibfnamefont {A.}~\bibnamefont {Devishvili}}, \bibinfo {author}
  {\bibfnamefont {V.}~\bibnamefont {Ukleev}}, \bibinfo {author} {\bibfnamefont
  {V.}~\bibnamefont {Kapaklis}}, \bibinfo {author} {\bibfnamefont
  {A.}~\bibnamefont {Oleshkevych}}, \bibinfo {author} {\bibfnamefont
  {V.}~\bibnamefont {Fedorov}}, \bibinfo {author} {\bibfnamefont
  {V.}~\bibnamefont {Bairamukov}}, \bibinfo {author} {\bibfnamefont
  {A.}~\bibnamefont {Vorobiev}},\ and\ \bibinfo {author} {\bibfnamefont
  {S.}~\bibnamefont {Grigoriev}},\ }\bibfield  {title} {\bibinfo {title}
  {Magnetic field induced chirality in ho/y multilayers with gradually
  decreasing anisotropy},\ }\bibfield  {journal} {\bibinfo  {journal} {Physical
  Review B}\ }\textbf {\bibinfo {volume} {96}},\ \href
  {https://doi.org/10.1103/physrevb.96.014415} {10.1103/physrevb.96.014415}
  (\bibinfo {year} {2017})\BibitemShut {NoStop}%
\bibitem [{\citenamefont {Gukasov}(1999)}]{Gukasov99}%
  \BibitemOpen
  \bibfield  {author} {\bibinfo {author} {\bibfnamefont {A.}~\bibnamefont
  {Gukasov}},\ }\bibfield  {title} {\bibinfo {title} {Left{\textendash}right
  asymmetry in polarised neutron scattering},\ }\href
  {https://doi.org/10.1016/s0921-4526(99)00007-1} {\bibfield  {journal}
  {\bibinfo  {journal} {Physica B: Condensed Matter}\ }\textbf {\bibinfo
  {volume} {267-268}},\ \bibinfo {pages} {97} (\bibinfo {year}
  {1999})}\BibitemShut {NoStop}%
\bibitem [{\citenamefont {Felcher}\ \emph {et~al.}(1995)\citenamefont
  {Felcher}, \citenamefont {Adenwalla}, \citenamefont {De~Haan},\ and\
  \citenamefont {Van~Well}}]{felcher95}%
  \BibitemOpen
  \bibfield  {author} {\bibinfo {author} {\bibfnamefont {G.}~\bibnamefont
  {Felcher}}, \bibinfo {author} {\bibfnamefont {S.}~\bibnamefont {Adenwalla}},
  \bibinfo {author} {\bibfnamefont {V.}~\bibnamefont {De~Haan}},\ and\ \bibinfo
  {author} {\bibfnamefont {A.}~\bibnamefont {Van~Well}},\ }\bibfield  {title}
  {\bibinfo {title} {Zeeman splitting of surface-scattered neutrons},\
  }\href@noop {} {\bibfield  {journal} {\bibinfo  {journal} {Nature}\ }\textbf
  {\bibinfo {volume} {377}},\ \bibinfo {pages} {409} (\bibinfo {year}
  {1995})}\BibitemShut {NoStop}%
\bibitem [{\citenamefont {Aksenov}\ \emph {et~al.}(2001)\citenamefont
  {Aksenov}, \citenamefont {Nikitenko},\ and\ \citenamefont
  {Kozhevnikov}}]{Aksenov01}%
  \BibitemOpen
  \bibfield  {author} {\bibinfo {author} {\bibfnamefont {V.}~\bibnamefont
  {Aksenov}}, \bibinfo {author} {\bibfnamefont {Y.}~\bibnamefont {Nikitenko}},\
  and\ \bibinfo {author} {\bibfnamefont {S.}~\bibnamefont {Kozhevnikov}},\
  }\bibfield  {title} {\bibinfo {title} {Spin-flip spatial neutron beam
  splitting in magnetic media},\ }\href
  {https://doi.org/10.1016/s0921-4526(00)00813-9} {\bibfield  {journal}
  {\bibinfo  {journal} {Physica B: Condensed Matter}\ }\textbf {\bibinfo
  {volume} {297}},\ \bibinfo {pages} {94} (\bibinfo {year} {2001})}\BibitemShut
  {NoStop}%
\bibitem [{\citenamefont {Kozhevnikov}\ \emph {et~al.}(2012)\citenamefont
  {Kozhevnikov}, \citenamefont {Ott},\ and\ \citenamefont
  {Radu}}]{Kozhevnikov12}%
  \BibitemOpen
  \bibfield  {author} {\bibinfo {author} {\bibfnamefont {S.}~\bibnamefont
  {Kozhevnikov}}, \bibinfo {author} {\bibfnamefont {F.}~\bibnamefont {Ott}},\
  and\ \bibinfo {author} {\bibfnamefont {F.}~\bibnamefont {Radu}},\ }\bibfield
  {title} {\bibinfo {title} {Data representations of zeeman spatial beam
  splitting in polarized neutron reflectometry},\ }\href
  {https://doi.org/10.1107/s0021889812018043} {\bibfield  {journal} {\bibinfo
  {journal} {Journal of Applied Crystallography}\ }\textbf {\bibinfo {volume}
  {45}},\ \bibinfo {pages} {814} (\bibinfo {year} {2012})}\BibitemShut
  {NoStop}%
\bibitem [{\citenamefont {Maranville}\ \emph {et~al.}(2016)\citenamefont
  {Maranville}, \citenamefont {Kirby}, \citenamefont {Grutter}, \citenamefont
  {Kienzle}, \citenamefont {Majkrzak}, \citenamefont {Liu},\ and\ \citenamefont
  {Dennis}}]{Maranville16}%
  \BibitemOpen
  \bibfield  {author} {\bibinfo {author} {\bibfnamefont {B.~B.}\ \bibnamefont
  {Maranville}}, \bibinfo {author} {\bibfnamefont {B.~J.}\ \bibnamefont
  {Kirby}}, \bibinfo {author} {\bibfnamefont {A.~J.}\ \bibnamefont {Grutter}},
  \bibinfo {author} {\bibfnamefont {P.~A.}\ \bibnamefont {Kienzle}}, \bibinfo
  {author} {\bibfnamefont {C.~F.}\ \bibnamefont {Majkrzak}}, \bibinfo {author}
  {\bibfnamefont {Y.}~\bibnamefont {Liu}},\ and\ \bibinfo {author}
  {\bibfnamefont {C.~L.}\ \bibnamefont {Dennis}},\ }\bibfield  {title}
  {\bibinfo {title} {Measurement and modeling of polarized specular neutron
  reflectivity in large magnetic fields},\ }\href
  {https://doi.org/10.1107/s1600576716007135} {\bibfield  {journal} {\bibinfo
  {journal} {Journal of Applied Crystallography}\ }\textbf {\bibinfo {volume}
  {49}},\ \bibinfo {pages} {1121} (\bibinfo {year} {2016})}\BibitemShut
  {NoStop}%
\bibitem [{\citenamefont {Kozhevnikov}\ \emph {et~al.}(2017)\citenamefont
  {Kozhevnikov}, \citenamefont {Ott},\ and\ \citenamefont
  {Semenova}}]{Kozhevnikov17}%
  \BibitemOpen
  \bibfield  {author} {\bibinfo {author} {\bibfnamefont {S.}~\bibnamefont
  {Kozhevnikov}}, \bibinfo {author} {\bibfnamefont {F.}~\bibnamefont {Ott}},\
  and\ \bibinfo {author} {\bibfnamefont {E.}~\bibnamefont {Semenova}},\
  }\bibfield  {title} {\bibinfo {title} {Neutron zeeman beam-splitting for the
  investigation of magnetic nanostructures},\ }\href
  {https://doi.org/10.1016/j.physb.2016.12.015} {\bibfield  {journal} {\bibinfo
   {journal} {Physica B: Condensed Matter}\ }\textbf {\bibinfo {volume}
  {508}},\ \bibinfo {pages} {12} (\bibinfo {year} {2017})}\BibitemShut
  {NoStop}%
\bibitem [{\citenamefont {Kozhevnikov}\ \emph {et~al.}(2018)\citenamefont
  {Kozhevnikov}, \citenamefont {Ignatovich},\ and\ \citenamefont
  {Radu}}]{Kozhevnikov18}%
  \BibitemOpen
  \bibfield  {author} {\bibinfo {author} {\bibfnamefont {S.~V.}\ \bibnamefont
  {Kozhevnikov}}, \bibinfo {author} {\bibfnamefont {V.~K.}\ \bibnamefont
  {Ignatovich}},\ and\ \bibinfo {author} {\bibfnamefont {F.}~\bibnamefont
  {Radu}},\ }\bibfield  {title} {\bibinfo {title} {On the application of zeeman
  spatial beam splitting in polarized neutron reflectometry},\ }\href
  {https://doi.org/10.1134/s1027451018010275} {\bibfield  {journal} {\bibinfo
  {journal} {Journal of Surface Investigation: X-ray, Synchrotron and Neutron
  Techniques}\ }\textbf {\bibinfo {volume} {12}},\ \bibinfo {pages} {103}
  (\bibinfo {year} {2018})}\BibitemShut {NoStop}%
\bibitem [{\citenamefont {Chen}\ \emph {et~al.}(2013)\citenamefont {Chen},
  \citenamefont {Zhu}, \citenamefont {Quesada}, \citenamefont {Li},
  \citenamefont {N'Diaye}, \citenamefont {Huo}, \citenamefont {Ma},
  \citenamefont {Chen}, \citenamefont {Kwon}, \citenamefont {Won},
  \citenamefont {Qiu}, \citenamefont {Schmid},\ and\ \citenamefont
  {Wu}}]{Chen13}%
  \BibitemOpen
  \bibfield  {author} {\bibinfo {author} {\bibfnamefont {G.}~\bibnamefont
  {Chen}}, \bibinfo {author} {\bibfnamefont {J.}~\bibnamefont {Zhu}}, \bibinfo
  {author} {\bibfnamefont {A.}~\bibnamefont {Quesada}}, \bibinfo {author}
  {\bibfnamefont {J.}~\bibnamefont {Li}}, \bibinfo {author} {\bibfnamefont
  {A.~T.}\ \bibnamefont {N'Diaye}}, \bibinfo {author} {\bibfnamefont
  {Y.}~\bibnamefont {Huo}}, \bibinfo {author} {\bibfnamefont {T.~P.}\
  \bibnamefont {Ma}}, \bibinfo {author} {\bibfnamefont {Y.}~\bibnamefont
  {Chen}}, \bibinfo {author} {\bibfnamefont {H.~Y.}\ \bibnamefont {Kwon}},
  \bibinfo {author} {\bibfnamefont {C.}~\bibnamefont {Won}}, \bibinfo {author}
  {\bibfnamefont {Z.~Q.}\ \bibnamefont {Qiu}}, \bibinfo {author} {\bibfnamefont
  {A.~K.}\ \bibnamefont {Schmid}},\ and\ \bibinfo {author} {\bibfnamefont
  {Y.~Z.}\ \bibnamefont {Wu}},\ }\bibfield  {title} {\bibinfo {title} {Novel
  chiral magnetic domain wall structure {inFe}/ni/cu(001)films},\ }\bibfield
  {journal} {\bibinfo  {journal} {Physical Review Letters}\ }\textbf {\bibinfo
  {volume} {110}},\ \href {https://doi.org/10.1103/physrevlett.110.177204}
  {10.1103/physrevlett.110.177204} (\bibinfo {year} {2013})\BibitemShut
  {NoStop}%
\bibitem [{\citenamefont {Shahbazi}\ \emph {et~al.}(2018)\citenamefont
  {Shahbazi}, \citenamefont {Hrabec}, \citenamefont {Moretti}, \citenamefont
  {Ward}, \citenamefont {Moore}, \citenamefont {Jeudy}, \citenamefont
  {Martinez},\ and\ \citenamefont {Marrows}}]{Shahbazi18}%
  \BibitemOpen
  \bibfield  {author} {\bibinfo {author} {\bibfnamefont {K.}~\bibnamefont
  {Shahbazi}}, \bibinfo {author} {\bibfnamefont {A.}~\bibnamefont {Hrabec}},
  \bibinfo {author} {\bibfnamefont {S.}~\bibnamefont {Moretti}}, \bibinfo
  {author} {\bibfnamefont {M.~B.}\ \bibnamefont {Ward}}, \bibinfo {author}
  {\bibfnamefont {T.~A.}\ \bibnamefont {Moore}}, \bibinfo {author}
  {\bibfnamefont {V.}~\bibnamefont {Jeudy}}, \bibinfo {author} {\bibfnamefont
  {E.}~\bibnamefont {Martinez}},\ and\ \bibinfo {author} {\bibfnamefont
  {C.~H.}\ \bibnamefont {Marrows}},\ }\bibfield  {title} {\bibinfo {title}
  {Magnetic properties and field-driven dynamics of chiral domain walls in
  epitaxial pt/co/{AuxPt}1-x trilayers},\ }\bibfield  {journal} {\bibinfo
  {journal} {Physical Review B}\ }\textbf {\bibinfo {volume} {98}},\ \href
  {https://doi.org/10.1103/physrevb.98.214413} {10.1103/physrevb.98.214413}
  (\bibinfo {year} {2018})\BibitemShut {NoStop}%
\bibitem [{\citenamefont {Ma}\ \emph {et~al.}(2017)\citenamefont {Ma},
  \citenamefont {Yu}, \citenamefont {Razavi}, \citenamefont {Sasaki},
  \citenamefont {Li}, \citenamefont {Hao}, \citenamefont {Tolbert},
  \citenamefont {Wang},\ and\ \citenamefont {Li}}]{Ma17}%
  \BibitemOpen
  \bibfield  {author} {\bibinfo {author} {\bibfnamefont {X.}~\bibnamefont
  {Ma}}, \bibinfo {author} {\bibfnamefont {G.}~\bibnamefont {Yu}}, \bibinfo
  {author} {\bibfnamefont {S.~A.}\ \bibnamefont {Razavi}}, \bibinfo {author}
  {\bibfnamefont {S.~S.}\ \bibnamefont {Sasaki}}, \bibinfo {author}
  {\bibfnamefont {X.}~\bibnamefont {Li}}, \bibinfo {author} {\bibfnamefont
  {K.}~\bibnamefont {Hao}}, \bibinfo {author} {\bibfnamefont {S.~H.}\
  \bibnamefont {Tolbert}}, \bibinfo {author} {\bibfnamefont {K.~L.}\
  \bibnamefont {Wang}},\ and\ \bibinfo {author} {\bibfnamefont
  {X.}~\bibnamefont {Li}},\ }\bibfield  {title} {\bibinfo {title}
  {Dzyaloshinskii-moriya interaction across an antiferromagnet-ferromagnet
  interface},\ }\bibfield  {journal} {\bibinfo  {journal} {Physical Review
  Letters}\ }\textbf {\bibinfo {volume} {119}},\ \href
  {https://doi.org/10.1103/physrevlett.119.027202}
  {10.1103/physrevlett.119.027202} (\bibinfo {year} {2017})\BibitemShut
  {NoStop}%
\end{thebibliography}%

\end{document}